\documentclass[preprint]{aastex}
\newcommand\pcc{\;{\rm cm}^{-3}}
\newcommand\Msun{\; {\rm M}_{\odot}}
\newcommand\kms{\; {\rm km}\;{\rm s}^{-1}}
\newcommand\ergs{\; {\rm erg}\;{\rm s}^{-1}}
\newcommand\erg{\; {\rm erg}}
\newcommand\mh{\; m_{\rm H}}

\newcommand\yr{\; {\rm yr}}
\newcommand\Myr{\;{\rm Myr}}
\newcommand\Gyr{\;{\rm Gyr}}
\newcommand\pc{\;{\rm pc}}
\newcommand\kpc{\;{\rm kpc}}
\newcommand\sfrunit{\Msun \kpc^{-2} \yr^{-1}}
\newcommand\Punit{\pcc\,{\rm K}}
\newcommand\Surf{\Msun\;{\rm pc^{-2}}}
\newcommand\rhounit{\Msun\;{\rm pc^{-3}}}

\newcommand\Kel{\;{\rm K}}

\newcommand\muG{\;\mu{\rm G}}

\newcommand\simgt{\lower.5ex\hbox{$\; \buildrel > \over \sim \;$}}
\newcommand\simlt{\lower.5ex\hbox{$\; \buildrel < \over \sim \;$}}

\newcommand\rbrackets[1]{\left({#1}\right)}

\newcommand\abrackets[1]{\left\langle{#1}\right\rangle}

\newcommand\vel{\mathbf{v}}

\newcommand\xhat{\hat{\mathbf{x}} }
\newcommand\yhat{\hat{\mathbf{y}} }
\newcommand\zhat{\hat{\mathbf{z}} }

\newcommand\rhosd{\rho_{\rm sd}}
\newcommand\kbol{k_{\rm B}}
\newcommand\Emagt{\delta E_{\rm mag}}
\newcommand\Emago{\overline{E}_{\rm mag}}
\newcommand\Eth{E_{\rm th}}
\newcommand\Ekin{E_{\rm turb}}
\newcommand\Pdriv{P_{\rm driv}}
\newcommand\Ptwo{P_{\rm two}}
\newcommand\Pmin{P_{\rm min}}
\newcommand\Pmax{P_{\rm max}}
\newcommand\Pth{P_{\rm th}}
\newcommand\Pturb{P_{\rm turb}}
\newcommand\Pmag{\Pi_{\rm mag}}
\newcommand\Pmagt{\delta \Pi_{\rm mag}}
\newcommand\Pmago{\overline{\Pi}_{\rm mag}}
\newcommand\Ptot{P_{\rm tot}}
\newcommand\PDE{P_{\rm tot,DE}}
\newcommand\Pfb{P_{\rm fb}}
\newcommand\etaturb{\eta_{\rm turb}}
\newcommand\etath{\eta_{\rm th}}

\newcommand\etamagt{\delta\eta_{\rm mag}}
\newcommand\etamago{\overline\eta_{\rm mag}}
\newcommand\vrms[2][1/2]{\langle #2^2\rangle_M^{#1}}
\newcommand\vth{v_{\rm th}}
\newcommand\rhomid{n_{\rm H,0}}
\newcommand\zmax{z_{\rm max}}
\newcommand\torb{t_{\rm orb}}

\newcommand\Wsg{\mathcal{W}_{\rm sg}}
\newcommand\Wext{\mathcal{W}_{\rm ext}}
\newcommand\SigSFR{\Sigma_{\rm SFR}}
\newcommand\SigSFRnorm{\Sigma_{\rm SFR,-3}}
\newcommand\gext{g_{\rm ext}}
\newcommand\sigmaeff{\sigma_\mathrm{eff}}

\shorttitle{Regulation of SFR in MHD simulations}
\shortauthors{Kim and Ostriker}

\begin{document}

\title{Vertical Equilibrium, Energetics, and Star Formation Rates in
Magnetized Galactic Disks Regulated by Momentum Feedback from Supernovae}

\author{Chang-Goo Kim and Eve C. Ostriker}

\affil{Department of Astrophysical Sciences, Princeton University, Princeton, NJ 08544, USA}
\email{cgkim@astro.princeton.edu, eco@astro.princeton.edu}
\slugcomment{\today}

\begin{abstract}
Recent hydrodynamic (HD) simulations have shown that galactic disks 
evolve to reach well-defined statistical equilibrium states.
The star formation rate (SFR) self-regulates until
energy injection by star formation feedback balances 
dissipation and cooling in the interstellar medium (ISM),
and provides vertical pressure support to balance gravity.
In this paper, we extend our previous models to allow for a range of initial magnetic field strengths and configurations,
utilizing three-dimensional, magnetohydrodynamic (MHD) simulations.
We show that a quasi-steady equilibrium state is established 
as rapidly for MHD as for HD models unless the initial 
magnetic field is very strong or very weak, which requires 
more time to reach saturation.  
Remarkably, models with initial magnetic 
energy varying by two orders of magnitude approach the same 
asymptotic state.
In the fully saturated state of the fiducial model, 
the integrated energy proportions 
$\Ekin:\Eth:\Emagt:\Emago$ are 
$0.35:0.39:0.15:0.11$, 
while the proportions of midplane support
$\Pturb:\Pth:\Pmagt:\Pmago$ are
$0.49:0.18:0.18:0.15$.
Vertical profiles of total effective pressure
satisfy vertical dynamical equilibrium with the 
total gas weight at all heights.
We measure the ``feedback yields'' $\eta_c\equiv P_c/\SigSFR$ 
(in suitable units) for 
each pressure component, finding that $\etaturb\sim 4$ and 
$\etath\sim 1$ are the same for MHD as in previous HD simulations, 
and $\etamagt \sim 1$. 
These yields can be used to predict the equilibrium SFR for a
local region in a galaxy based on its observed gas and stellar 
surface densities and velocity dispersions.
As the ISM weight (or dynamical equilibrium pressure) is fixed, an 
increase in $\eta$ from  
turbulent magnetic fields reduces the predicted $\SigSFR$ by 
$\sim 25\%$ relative to the HD case.  
\end{abstract}


\section{Introduction}\label{sec:intro}
The diffuse atomic interstellar medium (ISM), the mass reservoir in galaxies
from which molecular clouds and stars are born, is highly turbulent
\citep[e.g.,][]{2004ARA&A..42..211E} and strongly magnetized
\citep[e.g.,][]{2001SSRv...99..243B}.  The roles of turbulence and magnetic
fields in the context of star formation have been emphasized many times
\citep[see][for
reviews]{1987ARA&A..25...23S,2004RvMP...76..125M,2007ARA&A..45..565M}.
Additionally, the importance of turbulence and magnetic fields on large scales
is clear from comparison of each component's pressure  to the vertical weight
of the ISM disk  \citep[e.g.][]{1990ApJ...365..544B,2001RvMP...73.1031F}.  In
the Solar neighborhood, representing fairly typical conditions among star
forming galactic disks, turbulent and magnetic pressures are
roughly comparable to each
other \citep[e.g.,][]{2005ApJ...624..773H}, and about three times larger than
the thermal pressure in diffuse gas  \citep[e.g.,][]{2011ApJ...734...65J}.
Therefore, turbulence and magnetic fields provide most of the vertical support
to the gas layer, and are essential to include in models of the vertical
structure in galactic disks.  Moreover, the magnetic field has a significant
random component \citep[see][for recent review]{2015ASSL..407..483H}, which
presumably is related to observed turbulent velocities. This implies an
intriguing additional dimension to models in which the star formation rate
(SFR) is connected to the vertical dynamical equilibrium of the ISM disk via
turbulence driven by star formation feedback
\citep{2010ApJ...721..975O,2011ApJ...731...41O,2011ApJ...743...25K}.

Another important characteristic of the diffuse atomic ISM is its multiphase
nature, which emerges naturally as a consequence of thermal instability with a
realistic treatment of cooling and heating \citep{1965ApJ...142..531F}.  The
temperature and density of the cold and warm medium differ by about two orders
of magnitude
\citep{1969ApJ...155L.149F,1995ApJ...443..152W,2003ApJ...587..278W}.  The
inhomogeneous, multiphase structure of the diffuse atomic ISM is observed to be
ubiquitous
\citep[e.g.,][]{2003ApJ...586.1067H,2009ApJ...693.1250D,2013MNRAS.436.2352R,2015arXiv150301108M}.
In addition to the warm and cold phases, the hot phase that is created by
supernovae (SNe) is an essential ISM  component
\citep{1974ApJ...189L.105C,1977ApJ...218..148M,2001RvMP...73.1031F}.  This
phase contains very little mass, but is of great dynamical importance because
the expansion of high-pressure SN remnants and superbubbles is the main driver
of turbulence in the surrounding denser phases of the ISM (and also drives
galactic winds)
\citep{2004A&A...425..899D,2006ApJ...653.1266J,2009ApJ...704..137J,2012ApJ...750..104H,2013MNRAS.429.1922C,2015arXiv150607180L}.

Owing to the huge difference between cooling and dynamical times and
correspondingly large range of spatial scales, directly modeling the multiphase
ISM is numerically quite challenging.  However, the combination of increasing
computational resources, with robust algorithms for evolving gas dynamics and
thermodynamics simultaneously, have enabled development of highly sophisticated
ISM models in the last decade.    Direct numerical simulations of vertically
stratified galactic disks with SN driven turbulence and stiff source terms to
follow heating and cooling include those of
\citet{2004A&A...425..899D,2005A&A...436..585D,2006ApJ...653.1266J,
2009ApJ...704..137J,2012ApJ...750..104H,2013MNRAS.432.1396G,
2011ApJ...743...25K,2013ApJ...776....1K,2014A&A...570A..81H,2014arXiv1412.2749W}.
Among those, only \citet[][hereafter Paper~I]{2013ApJ...776....1K} and
\citet{2014A&A...570A..81H} have a time-dependent SFR and SN rate
self-consistently set by the self-gravitating localized collapse of gas.  

The time-dependent response of the SFR to large-scale ISM dynamics and
thermodynamics is key to self-consistent regulation of turbulent and thermal
pressures, since both the turbulent driving rate (from SNe) and thermal heating
rate (from far-UV-emitting massive stars) are proportional to the SFR. If
heating rates and turbulent driving rates drop due to a low SFR, the increased
mass of cold, dense gas subsequently leads to a rebound in the SFR.  The
opposite is also true, with an excessively high SFR checked by the ensuing
reduction in the mass of gas eligible to collapse.  Within less than an orbital
time, the SFR self-regulates such that the ISM achieves a
 statistical equilibrium state.
In this state, there are balances between (1) vertical gravity and the combined
pressure forces, (2) turbulent driving and dissipation, and (3) cooling and
heating (see
\citealt{2010ApJ...721..975O,2011ApJ...731...41O,2011ApJ...743...25K} for
analytic theory, and
\citealt[][Paper~I]{2011ApJ...743...25K,2012ApJ...754....2S} for numerical
confirmations).  Paper~I successfully reproduces important aspects of the
diffuse atomic ISM, demonstrating self-regulated SFR and vertical equilibrium
following the analytic theory for a range of galactic conditions including
those of the Solar neighborhood as well as regions with higher and lower total
ISM surface density.  In addition, the warm/cold ratios and properties of 21 cm
emission/absorption for  \ion{H}{1}  (including column density and spin
temperature distributions) are in agreement with observations
\citep{2014ApJ...786...64K}.

Despite of the success of hydrodynamic (HD) models in Paper~I, the dynamical
role of magnetic fields cannot be ignored.  For example,
\citet{2014A&A...570A..81H} have run magnetized ISM disk models with
self-gravity and SN feedback, showing a factor of two decrement in the SFR
compared to the HD case. However, a systematic exploration of the effect of
magnetic fields is still lacking.  In this paper, a set of three-dimensional,
magnetohydrodynamic (MHD) simulations is carried out, to test the effects of
different initial magnetic field strengths and configuration.

The remainder of the paper is organized as follows.  In
Section~\ref{sec:theory}, we begin by reviewing the analytic theory and
previous simulation results to identify and explain the properties and
parameters we will measure here.  Section~\ref{sec:method} summarizes our
numerical methods and model parameters. Our basis is the fiducial Solar
neighborhood model from Paper~I with gas surface density of 10$\Surf$, but here
we slightly alter the initial and boundary conditions to minimize early
vertical oscillations that were exaggerated in Paper~I.
Section~\ref{sec:results} contains our main new results.  Time evolution of
disk diagnostics in Section~\ref{sec:evol} shows that a quasi-steady state is
achieved.  In Section~\ref{sec:equil}, we analyze temporally and horizontally
averaged vertical profiles to confirm vertical dynamical equilibrium.  Finally,
Section~\ref{sec:sfr} presents the measurements of ``feedback yields,''
exploring the mutual relationship between measured SFR surface density and the
various midplane pressure supports. We compare the numerical results with
analytic expectations including magnetic terms.  In Section~\ref{sec:summary},
we summarize and discuss the model disk properties at saturation in comparison
to observations and previous simulations, especially focusing on the level of
mean and turbulent magnetic fields.

\section{The Equilibrium Theory}\label{sec:theory}

Before describing the results of our new simulations, it is informative to
summarize the analytic theory for mutual equilibrium of the ISM and star
formation in disk systems, as developed in our previous work
\citep{2010ApJ...721..975O,2011ApJ...731...41O,2011ApJ...743...25K}.  This
includes a list of the parameters to be measured for detailed investigation.
The equilibrium theory assumes a quasi-steady state that satisfies force
balance between vertical gravity and pressure support, and energy balance
between gains from star formation feedback and losses in the dissipative ISM.
We shall test the validity of these assumptions and quantify the relative
contributions to the vertical support from turbulent, thermal, and magnetic
terms. We shall also measure the efficiency of feedback (feedback yield) for
each support term.

Using the temporally- and horizontally- averaged momentum equation, it is
straightforward to show that vertical dynamical equilibrium requires a balance
between the total momentum flux difference between surfaces at $z=0$ and
$\zmax$, and the weight of the gas: $\Delta \Ptot=\mathcal{W}$.  In magnetized,
turbulent galactic disks, the vertical momentum flux consists of thermal
($\Pth\equiv\rho \vth^2$), turbulent ($\Pturb\equiv\rho v_z^2$), and magnetic
($\Pmag\equiv|\mathbf{B}|^2/8\pi-B_z^2/4\pi$) terms
\citep[e.g.,][]{1990ApJ...365..544B,2007ApJ...663..183P,2011ApJ...731...41O}\footnote{
More generally, radiation and cosmic ray pressure terms could also contribute
\citep[see][]{2011ApJ...731...41O}, but are not included in the present
numerical simulations}.  Note that the magnetic term includes both pressure and
tension, and can be rewritten as $\Pmag=\rho(v_A^2/2-v_{A,z}^2)$,
where $v_A\equiv \left(|\mathbf{B}|^2/4\pi\rho\right)^{1/2}$ is the Alfv\'en
velocity from the total magentic field and $v_{A,z}=|B_z|/(4\pi \rho)^{1/2}$ is
from its $z$ component.  We hereafter refer to $\Pmag$ as the magnetic
``support'' to distinguish from the usual magnetic ``pressure'' ($P_{\rm
mag}\equiv|\mathbf{B}|^2/8\pi$), while the thermal and turbulent ``supports''
are equivalent to the thermal and turbulent ``pressures,'' respectively.  If we
choose $\zmax$ where the gas density is sufficiently small, $\Pth(\zmax)$,
$\Pturb(\zmax)\rightarrow 0$ by definition, but $\Pmag(\zmax)$ can in general
be nonzero and significant. We thus have
\begin{equation}\label{eq:Ptot}
\Delta \Ptot= {\Pth}_{,0}+{\Pturb}_{,0}+\Delta\Pmag
\equiv \rho_0 \sigma_z^2(1+\mathcal{R}),
\end{equation}
where $\sigma_z^2\equiv \vth^2+v_z^2$ is the sum of thermal and turbulent
velocity dispersions, and
\begin{equation}\label{eq:R}
\mathcal{R}\equiv \frac{\Delta\Pmag}{\rho_0 \sigma_z^2}
\end{equation}
is the relative contribution to the vertical support from magnetic to kinetic
(thermal plus turbulent) terms. If $\mathbf{B}(\zmax)\rightarrow 0$, $\Delta
\Pmag \rightarrow (|\mathbf{B}_0|^2/2 -B_{z,0}^2)/4\pi$.  Here and hereafter,
we use the subscript `0' to indicate quantities evaluated at the midplane
($z=0$),

The weight of the gas under self- and external gravity is respectively defined
by
\begin{equation}\label{eq:wsg}
\Wsg = \int_0^{\zmax} \rho \frac{d\Phi_{\rm sg}}{dz} dz = \frac{\pi G\Sigma^2}{2}
\end{equation}
and
\begin{equation}\label{eq:wext}
\Wext = \int_0^{\zmax} \rho \gext(z) dz \equiv \zeta_d\gext(H)\Sigma
\end{equation}
Here, $\Sigma\equiv \int_{-\infty}^\infty \rho dz$ is the gaseous surface
density, $H\equiv \Sigma/(2\rho_0)$ is the gaseous disk's effective scale
height, and $\zeta_d\equiv (1/2)\int_{0}^{\zmax} \rho
|\gext(z)/\gext(H)|dz/\int_{0}^{\zmax} \rho dz$ is a dimensionless parameter
that characterizes the vertical distribution of gas and the shape of external
gravity profile.  For example, with linear external gravity profile
($\gext\propto z$), density following exponential, Gaussian, and sech$^2$
distributions gives $\zeta_d=1/2$, $1/\pi$, and $(\ln2)/2$, respectively. 

We now specialize to the case of a  galactic disk with a midplane density of
stars + dark matter equal to $\rhosd$, for which the external gravity near the
midplane is $g_\mathrm{ext}=4 \pi G \rhosd z$.  Vertical dynamical equilibrium
can be expressed as
\begin{equation}\label{eq:de}
\rho_0 \sigma_z^2(1+\mathcal{R})=\Wsg(1+\chi),
\end{equation}
where 
\begin{equation}\label{eq:chidef}
\chi\equiv \frac{\Wext}{\Wsg} =\frac{4\zeta_d\rhosd}{\rho_0}
\end{equation}
is the ratio of external to self gravity.  By substituting from Equation
(\ref{eq:wsg}) for $\Wsg$ and from Equation (\ref{eq:chidef}) for $\chi$ in
Equation (\ref{eq:de}), we can solve to obtain 
\begin{equation}\label{eq:H}
H = H_{\rm sg}\frac{1+\mathcal{R}}{1+\chi}
= H_{\rm ext}\rbrackets{\frac{1+\mathcal{R}}{1+1/\chi}}^{1/2},
\end{equation}
where $H_{\rm sg}\equiv\sigma_z^2/\pi G\Sigma$ and $H_{\rm ext}\equiv
\sigma_z/(8\pi G\zeta_d\rhosd)^{1/2}$ are the scale heights for self- and
external- gravity dominated cases, respectively.  By defining 
\begin{equation}\label{eq:Cdef}
C\equiv \frac{H_{\rm sg}^2}{H_{\rm ext}^2}=\frac{8\zeta_d \rhosd \sigma_z^2}{\pi G\Sigma^2},
\end{equation}
Equation (\ref{eq:H}) can also be solved for  $\chi$ to obtain
\begin{equation}\label{eq:chi}
\chi=\frac{2C(1+\mathcal{R})}{1+\sqrt{1+4C(1+\mathcal{R})}}.
\end{equation}
If we neglect dark matter and consider a stellar disk with surface density
$\Sigma_*$ and vertical stellar velocity dispersion $\sigma_*$ such that
$\rhosd \rightarrow \pi G \Sigma_*^2/(2\sigma_*^2)$, we have
\begin{equation}\label{eq:C}
C(1+\mathcal{R})=4\zeta_d \rbrackets{\frac{\sigmaeff\Sigma_*}{\sigma_*\Sigma}}^2.
\end{equation}
Here,
\begin{equation}\label{eq:sigmaeff}
\sigmaeff\equiv\left(\vth^2+v_z^2+\frac{v_A^2}{2}-v_{A,z}^2\right)^{1/2}=\sigma_z(1+ \mathcal{R})^{1/2}
\end{equation}
is the effective vertical velocity dispersion including magnetic terms for
gaseous support, and we have assumed that $\mathbf{B}(\zmax)\rightarrow 0$.

We shall show from results of our simulations in Section~\ref{sec:evol} that
$\sigmaeff\sim 5-6\kms$ and $\zeta_d\sim 0.4-0.5$ are quite insensitive to the
initial magnetization (see Tables~\ref{tbl:mean} and \ref{tbl:meanv}).  In
Paper~I, we also found relatively little variation in $\sigma_z$, even with a
factor 10 variation in $\Sigma$ and two orders of magnitude variation  of
$\rhosd/\Sigma^2$, the input parameter that controls the ratio of external- to
self-gravity.  In this paper, we fix $\Sigma$ and $\rhosd$ to isolate the
effect of magnetic fields. Further investigations would therefore be needed to
investigate constancy or variation of $\sigmaeff$ and $\zeta_d$ with $\Sigma$
and $\rhosd$, which we defer to future work.

In simulations, $\Sigma$ and $\rhosd$ (as well as a seed magnetic field) are
input parameters, and $\sigma_z$, $\mathcal R$, and $\zeta_d$ are measured
outputs.  From the right-hand side of Equation (\ref{eq:de}) combined with
Equations (\ref{eq:chi}) and (\ref{eq:Cdef}), the predicted dynamical
equilibrium midplane pressure can be obtained.  In observed (relatively
face-on) galactic disks, the gas and stellar surface densities $\Sigma$ and
$\Sigma_*$ together with $\sigma_z$ and $\sigma_*$  may be considered basic
observables \citep[although the stellar scale height $H_* = \sigma_*^2/(\pi G
\Sigma_*)$ instead may be estimated in other ways if $\sigma_*$ is not directly
measured; e.g. ][]{2008AJ....136.2782L}. The magnetic term $v_A^2/2- v_{A,z}^2$
in $\sigmaeff^2$ is more difficult to measure directly.  We shall show,
however, that if the model disk is fully saturated, $\mathcal{R}=(v_A^2/2-
v_{A,z}^2)/\sigma_z^2$  appears to be insensitive to the initial magnetic
geometry or strength.  This implies that the predicted equilibrium midplane
pressure support may similarly be obtained from observables using Equations
(\ref{eq:de}), (\ref{eq:chi}), (\ref{eq:C}), (\ref{eq:sigmaeff}) if estimates
of $\mathcal{R}$ and $\zeta_d$ from simulations are adopted.

To test the validity of vertical dynamical equilibrium, for our simulations we
will compare full vertical profiles of the total (turbulent, thermal, and
magnetic) support with the vertical profiles of the weight of the gas.  These
profiles are based on horizontal and temporal averages of the simulation
outputs at all values of $z$, including the midplane. 
In equilibrium, the weight and total support profiles must match each other.   

Due to the highly dissipative nature of the ISM, continuous energy injection is
necessary to maintain thermal as well as turbulent kinetic and magnetic
components of the vertical support at a given level.  Massive young stars
inject prodigious energy, providing the feedback that is key to self-regulation
of the SFR.  When the system is out of equilibrium, with not enough massive
young stars, lack of energy injection leads the entire ISM to become
dynamically and thermally cold.  A cold disk is highly susceptible to
gravitational collapse; it forms new stars that supply the ``missing" feedback
and restore equilibrium.  In the opposite case, with too many massive stars,
the ISM becomes dynamically and thermally hot, quenching further star formation
by suppressing gravitational instability.  Simulations of local model disks
show that the SFR converges to the quasi-steady value predicted by theory, in
which the vertical support produced by feedback matches the requirements set by
vertical dynamical equilibrium.  The ISM state and SFR predicted by the
feedback-regulated theory are expected to hold in real galaxies provided all
equilibria can be established in  less than  the disk's secular evolution
timescale. 

In Paper~I \citep[see also][]{2011ApJ...743...25K,2012ApJ...754....2S}, we
showed that the balance between energy gains and losses is established within
one vertical crossing time, the turbulence dissipation timescale. We then
quantified the efficiency of energy conversion to each support component for a
given SFR by measuring the ``feedback yield.''

Using a suitable normalization\footnote{
Dimensionally, $\eta$ has units of velocity.  To obtain $\eta$ in
$\kms$, the values reported in this paper should be multiplied by 209.}, 
we define yield parameters $\eta_c$ as
\begin{equation}\label{eq:eta}
\eta_c\equiv\frac{P_{c,3}}{\SigSFRnorm}
\end{equation}
where $P_{c,3}=P_{c,0}/10^3\kbol\Punit$ and
$\SigSFRnorm=\SigSFR/10^{-3}\sfrunit$.  The subscript `$c$' denotes ``turb,''
``th," or ``mag" for the respective  component of vertical support  ($\Pturb$,
$\Pth$, or $\Pmag$).  For magnetic pressure, the support is further divided
into turbulent and mean components, $\Pmagt$ and $\Pmago$, respectively (see
Section~\ref{sec:evol} for definitions).  \citet{2010ApJ...721..975O} and
\citet{2011ApJ...731...41O} respectively showed that $\etath$ and $\etaturb$
are expected to be nearly independent of $\SigSFR$.  In Paper~I \citep[see
also][]{2011ApJ...743...25K}, which omitted magnetic fields but covered a wide
range of disk conditions such that $0.1<\SigSFRnorm<10$, we obtained
$\etaturb=4.3\SigSFRnorm^{-0.11}$ and $\etath=1.3\SigSFRnorm^{-0.14}$. The
values (and weak $\SigSFR$ dependence) of these numerically calibrated yield
parameters are in very good agreement with analytic expectations. 

The MHD simulations of this paper allow us to study the evolution and
saturated-state properties of magnetic fields in star-forming, turbulent,
differentially-rotating galactic disks with vertical stratification. ISM
turbulence driven by star formation feedback can generate and deform magnetic
fields.  The small scale turbulent dynamo, combined with buoyancy and sheared
rotation, creates turbulent magnetic fields  and modifies mean magnetic fields,
both of which can provide vertical support to the ISM.  We shall consider three
different initial magnetic field strengths (as well as two initial vertical
profiles), and directly measure the saturated-state magnetic field strengths
and feedback yields, while also testing how magnetization affects the thermal
and turbulent feedback yield components.

\section{Methods and Models}\label{sec:method}

In this paper, we extend our previous three dimensional HD simulations from
Paper~I to include magnetic fields.  We solve ideal MHD equations in a local,
shearing box with self- and external gravity, thermal conduction, 
and optically thin cooling.
To solve the MHD equations, we utilize \emph{Athena}
\citep{2008ApJS..178..137S} with the van Leer integrator
\citep{2009NewA...14..139S}, HLLD solver, and second-order spatial
reconstruction. 
We also consider feedback from massive young stars 
using time-varying heating rate and momentum feedback from SNe.
The probability of massive star formation is calculated
based on a predicted local SFR ($\dot{M}_*$)
in cells with density exceeding a threshold, assuming an efficiency per
free-fall time $\epsilon_{\rm ff}=1\%$ \citep[e.g.,][]{2007ApJ...654..304K},
and adopting a total mass in new stars per massive star of $m_*=100\Msun$.
When a massive star forms (i.e., when a uniform random number in (0,1)
is less than the probability $\dot{M}_*/m_*\Delta t$), 
we immediately inject total radial momentum $p_*=3\times10^5 \Msun\kms$ in a
surrounding $10\pc$ sphere. 
The global SFR surface density is then calculated by 
\begin{equation}\label{eq:sfr}
\SigSFR = \frac{\mathcal{N}_{*} m_*}{L_xL_y t_{\rm bin}},
\end{equation}
where $\mathcal{N}_{*}$ denotes the total number of massive stars in the time
interval $(t-t_{\rm bin},t)$, and $t_{\rm bin}=10\Myr$ is UV-weighted lifetime
of OB stars \citep[e.g.,][]{2003ApJ...584..797P}.
The heating rate is proportional to the mean FUV intensity, 
which is assumed to be linearly proportional to the SFR surface density,
$\Gamma\propto J_{\rm FUV}\propto \SigSFR$.
See Paper~I for more details and other source terms.

The models in this paper are magnetically-modified versions of the fiducial
simulation QA10 from Paper~I, with conditions similar to the Solar
neighborhood. Our simulation domain is a local Cartesian grid far from the
galactic center, with center rotating at angular velocity of
$\Omega=28\kms\kpc^{-1}$ (the corresponding orbital time is
$\torb=2\pi/\Omega=219\Myr$) and shear parameter $q\equiv-d\ln\Omega/d\ln R=1$
for a flat rotation curve.  In the $\hat z$ direction, we adopt outflow
boundary conditions to allow magnetic flux loss at the vertical boundaries;
this is in contrast to Paper~I, where we adopted periodic vertical boundary
conditions (except for the gravitational potential) to prevent any mass loss.
In order to minimize mass loss at the vertical boundaries, we double the
vertical domain size to $L_z=1024\pc$ compared to Paper~I. Shearing-periodic
boundary conditions are employed in the horizontal directions, with
$L_x=L_y=512\pc$ the same as in Paper~I.  The spatial resolution is set to
$2\pc$ as in Paper~I.  We adopt a linear external gravity profile
$\mathbf{g}_{\rm ext} = - 4\pi G\rhosd z \zhat$, where $\rhosd$ is the midplane
volume density of stellar disk plus dark matter.  

In order to explore the independent effect of the magnetic fields, we fix the
background gravity and gas surface density parameters to $\rhosd=0.05\rhounit$
and $\Sigma=10\Surf$ for all models.  Initial profiles of the gas density are
set by an exponential function as $\rho=\rho_0\exp(-|z|/H)$, with midplane
density  $\rho_0=\Sigma/2H$ and scale height set using $\sigma_z=4\kms$ in
Equations~(\ref{eq:H}) and (\ref{eq:Cdef}), also using $\zeta_d=1/2$ and
$\mathcal{R}$ as described below.  The initial  thermal pressure has the same
profile as the density with midplane thermal pressure of $P_{\rm
th,0}/\kbol=3000\Punit$.  We vary the plasma beta at the midplane
$\beta_0\equiv P_{\rm th,0}/P_{\rm mag,0} =8\pi P_{\rm th,0}/B_0^2$ with two
different vertical field distributions; uniform plasma beta with
$\mathbf{B}=B_0\exp(-|z|/2H)\yhat$ (Models MA), and uniform magnetic fields
with $\mathbf{B}=B_0\yhat$ (Models MB).  The suffixes of magnetized models
denote the values of $\beta_0=1$, 10, and 100.  The initial magnetic energy is
then largest for Model MB1 and smallest for Model MA100. For comparison, we
also present the results from an unmagnetized counterpart (Model HL) as well as
the previous Model QA10 from Paper~I (here, renamed HS).  These hydrodynamic
models differ in the size of the vertical domain (``S''=small, $L_z=512$ pc;
``L''=large, $L_z=1024$ pc).  

For our initial conditions, we have $\zeta_d=1/2$ and $C=2.37$ for all models.
With vertically stratified magnetic fields (Models MA;
$\mathcal{R}=1/\beta_0$), $\chi=1.12$, 1.13, 1.19, and 1.73 for
$\beta_0=\infty$, 100, 10, and 1, respectively, while $\chi=1.12$ for all MB
models ($\mathcal{R}=0$). The scale height is $H=77
[(1+\mathcal{R})(1+1/\chi)]^{1/2}\pc$ and
$\rhomid=\rho_0/1.4\mh=1.9[(1+\mathcal{R})(1+1/\chi)]^{-1/2}\pcc$. The midplane
magnetic field strength is
\begin{equation}
B_0=3.2\beta_0^{-1/2} 
\rbrackets{\frac{P_{\rm th,0}/\kbol}{3000\Punit}}^{1/2}\mu G .
\end{equation}

In contrast to Paper~I, we drive turbulence for $\torb$ in order to provide
turbulent support at early stages before SN feedback generates sufficient
turbulence.  We utilize divergence-free turbulent velocity fields following a
Gaussian random distribution with a power spectrum of the form $|\delta
\vel_k^2| \propto k^6 \exp(-8k/k_{\rm pk})$ where the peak driving is at
$k_{\rm pk}L_x/2\pi=4$ and $1\ge kL_x/2\pi\ge128$.  A new turbulent velocity
perturbation field is generated every $10\Myr$, with total  energy injection
rate of $\dot{E}_{\rm turb} = 500L_\odot$ to the turbulence.  This perturbation
corresponds to the saturation level of one-dimensional velocity dispersion
$\sim 4\kms$. Turbulence is driven at full strength up to $\torb/2$, and then
slowly turned off from $\torb/2$ to $\torb$.  The spatially uniform
photoelectric heating rate is set to constant for the first $\torb/2$
($\Gamma=0.8\Gamma_0$, where $\Gamma_0=2\times10^{-26}\ergs$ from
\citealt{2002ApJ...564L..97K}). From $\torb/2$ to $\torb$ this  imposed heating
is slowly turned off and replaced by the self-consistent heating rate that is
proportional to the SFR surface density ($\Gamma=0.4\Gamma_0\SigSFRnorm$).

By initially driving turbulence, and allowing smooth changes from early to
saturated stages in the thermal and turbulent supports, the current models
minimize abrupt early collapse that was seen in the models of Paper~I that were
initialized without turbulence (as reproduced in model HS here).  There, the
initial vertical collapse also triggered a strong burst of star formation,
leading to exaggerated vertical oscillations.  However, we shall show that
these and other differences in numerical treatment (including initial
turbulence driving, vertical boundary conditions, and domain size) make little
or no differences in the physical quantities (see Section~\ref{sec:evol}) and
averaged vertical distributions (see Section~\ref{sec:equil}). Since some
magnetized models saturate slowly, we run magnetized models longer to achieve a
quasi-steady state for the turbulent magnetic fields.  For the purpose of
computing time-averaged quantities, the saturated stages are considered from
$t_1=3\torb$ to $t_2=4\torb$ for magnetized models and from $t_1=1.5\torb$ to
$t_2=2\torb$ for unmagnetized models.  Note that 4 orbits is still not long
enough for complete saturation of the mean magnetic field in the cases of
initially strongest and weakest magnetization.

\section{Simulation Results}\label{sec:results}

\subsection{Time Evolution and Saturated State}\label{sec:evol}

In this section, we shall show that model disks achieve a quasi-steady state,
using evolution of diagnostics that describe the average disk properties at a
given time. Volume and mass-weighted means for quantities $q_{ijk}(t)$ are
respectively calculated by 
\begin{equation}\label{eq:mwavg}
\abrackets{q}_V(t)\equiv \frac{\sum q \Delta V}{L_x L_y L_z},
\quad
\abrackets{q}_M(t)\equiv \frac{\sum \rho q \Delta V}{\sum\rho \Delta V},
\end{equation}
where the summation is over all grid zones (indices $ijk$), and the volume
element is $\Delta V=(2\pc)^3$. We also calculate the horizontally-averaged
vertical profile as
\begin{equation}\label{eq:havg}
\overline{q}(z;t) \equiv \frac{\sum_{i,j} q \Delta x \Delta y}{L_xL_y},
\end{equation}
where the summation is only over horizontal planes (indices $ij$) at each
vertical coordinate $z$ (index $k$).

We use a horizontal average to obtain the mean magnetic field,
$\overline{\mathbf{B}}(z)$.  The turbulent magnetic field is then defined by
$\delta \mathbf{B} = \mathbf{B}-\overline{\mathbf{B}}$.  Note that in our
simulations the mean field is dominated by $\yhat$-component with small
$\xhat$-component and negligible $\zhat$-component.  The turbulent magnetic
pressure is given by $\delta P_\mathrm{mag} \equiv |\delta \mathbf{B}|^2/8\pi$
and the mean magnetic pressure is given by $\bar P_\mathrm{mag} \equiv
|\bar{\mathbf{B}}|^2/8\pi$.

\begin{figure}
\plotone{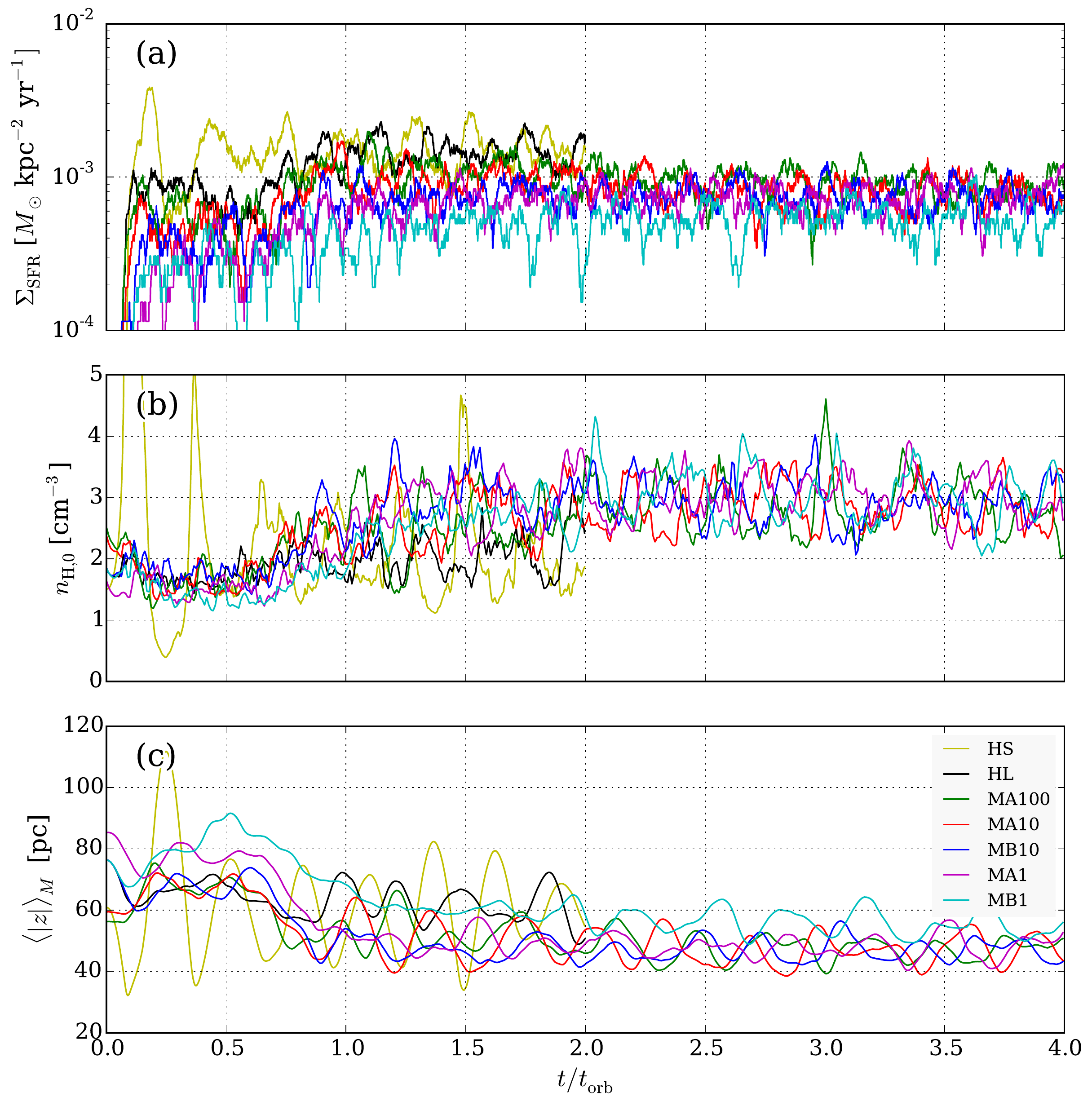}
\caption{Time evolution of
(a) SFR surface density $\SigSFR$, (b) midplane number density $\rhomid$, and
(c) mass-weighted mean thickness $\abrackets{|z|}_M$.  Physical model
parameters differ only in the initial magnetic energy, increasing from zero in
HS and HL to a  maximum in MB1 (see Section \ref{sec:method}).}
\label{fig:evol1}
\end{figure}

\begin{figure}
\plotone{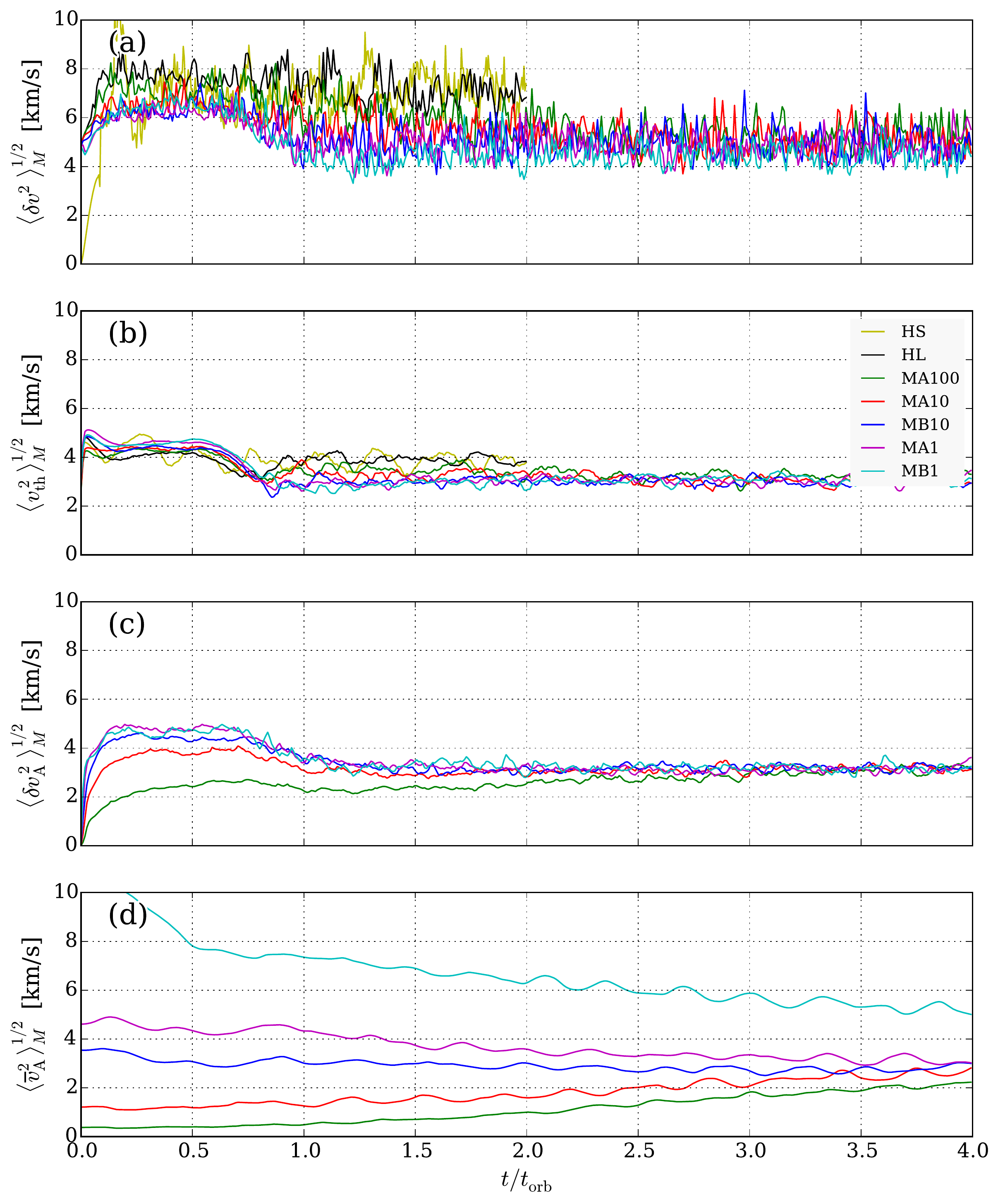}
\caption{Time evolution of mass-weighted RMS (a) turbulent kinetic, (b)
thermal, (c) turbulent Alfv\'en and (d) mean Alfv\'en velocity.  }
\label{fig:evol2}
\end{figure}

\begin{deluxetable}{lcccccccc}
\tablewidth{0pt}
\tabletypesize{\footnotesize}
\tablecaption{Mean Disk Properties at Saturation\label{tbl:mean}}
\tablehead{
\colhead{Model} &
\colhead{$\SigSFR$} &
\colhead{$\rhomid$} &
\colhead{$\langle{|z|}\rangle_M$} &
\colhead{$H$} &
\colhead{$\zeta_d$} &
\colhead{$\chi$} &
\colhead{$\alpha_{\rm ss,R}$} &
\colhead{$\alpha_{\rm ss,M}$}}
\startdata
HS &     $ 1.59 \pm  0.38$ &     $ 1.96 \pm  0.56$ &     $61.5 \pm 10.4$ &     $73.8 \pm 21.1$ &     $ 0.42 \pm  0.14$ &     $ 1.23 \pm  0.54$ &     $ 0.22 \pm  0.07$ &     \nodata \\
HL &     $ 1.46 \pm  0.28$ &     $ 2.15 \pm  0.41$ &     $60.6 \pm  6.3$ &     $67.3 \pm 12.9$ &     $ 0.45 \pm  0.10$ &     $ 1.21 \pm  0.35$ &     $ 0.25 \pm  0.10$ &     \nodata \\
MA100 &  $ 0.91 \pm  0.16$ &     $ 2.87 \pm  0.45$ &     $47.1 \pm  3.1$ &     $50.3 \pm  7.9$ &     $ 0.47 \pm  0.08$ &     $ 0.94 \pm  0.22$ &     $ 0.22 \pm  0.13$ &     $ 0.23 \pm  0.07$ \\
MA10 &   $ 0.81 \pm  0.16$ &     $ 2.87 \pm  0.35$ &     $47.4 \pm  4.2$ &     $50.4 \pm  6.2$ &     $ 0.47 \pm  0.07$ &     $ 0.95 \pm  0.18$ &     $ 0.22 \pm  0.09$ &     $ 0.27 \pm  0.07$ \\
MB10 &   $ 0.74 \pm  0.15$ &     $ 2.85 \pm  0.31$ &     $47.6 \pm  3.6$ &     $50.7 \pm  5.5$ &     $ 0.47 \pm  0.06$ &     $ 0.95 \pm  0.16$ &     $ 0.21 \pm  0.09$ &     $ 0.32 \pm  0.08$ \\
MA1 &    $ 0.75 \pm  0.16$ &     $ 3.00 \pm  0.37$ &     $48.2 \pm  4.2$ &     $48.2 \pm  6.0$ &     $ 0.50 \pm  0.08$ &     $ 0.96 \pm  0.19$ &     $ 0.23 \pm  0.10$ &     $ 0.22 \pm  0.10$ \\
MB1 &    $ 0.56 \pm  0.14$ &     $ 2.97 \pm  0.43$ &     $55.1 \pm  4.3$ &     $48.7 \pm  7.0$ &     $ 0.57 \pm  0.09$ &     $ 1.10 \pm  0.24$ &     $ 0.17 \pm  0.09$ &     $ 0.27 \pm  0.10$ \\
\enddata
\tablecomments{The mean and standard deviation are taken over
$t/\torb=1.5-2$ for HD models and $t/\torb=3-4$ for MHD models.
See Sections \ref{sec:theory} and \ref{sec:evol} for definition of each quantity.
$\SigSFR$ is in units of $10^{-3}\sfrunit$,
$\rhomid$ is in units of $\rm cm^{-3}$,
$\abrackets{|z|}_M$ and $H$ are in units of pc.}
\end{deluxetable}

\begin{deluxetable}{lcccccc}
\tablewidth{0pt}
\tabletypesize{\footnotesize}
\tablecaption{Saturated State Velocities\label{tbl:meanv}}
\tablehead{
\colhead{Model} &
\colhead{$\vrms{\delta v}$} &
\colhead{$\vrms{\vth}$} &
\colhead{$\vrms{\delta v_A}$} &
\colhead{$\vrms{\overline{v}_{A}}$} &
\colhead{$\sigma_{\rm z}$} &
\colhead{$\sigmaeff$}} 
\startdata
HS &	 $ 7.31 \pm  0.65$ &	 $ 3.89 \pm  0.20$ &	 \nodata           &	 \nodata           &	 $ 5.96 \pm  0.36$ &	 $ 5.96 \pm  0.36$ \\ 
HL &	 $ 6.95 \pm  0.53$ &	 $ 3.89 \pm  0.16$ &	 \nodata           &	 \nodata           &	 $ 5.67 \pm  0.27$ &	 $ 5.67 \pm  0.27$ \\ 
MA100 &	 $ 5.22 \pm  0.55$ &	 $ 3.16 \pm  0.13$ &	 $ 3.07 \pm  0.13$ &	 $ 1.93 \pm  0.17$ &	 $ 4.36 \pm  0.28$ &	 $ 4.78 \pm  0.27$ \\ 
MA10 &	 $ 5.10 \pm  0.55$ &	 $ 3.07 \pm  0.16$ &	 $ 3.18 \pm  0.10$ &	 $ 2.47 \pm  0.17$ &	 $ 4.27 \pm  0.29$ &	 $ 4.83 \pm  0.25$ \\ 
MB10 &	 $ 4.89 \pm  0.55$ &	 $ 3.01 \pm  0.15$ &	 $ 3.19 \pm  0.12$ &	 $ 2.74 \pm  0.13$ &	 $ 4.17 \pm  0.30$ &	 $ 4.81 \pm  0.27$ \\ 
MA1 &	 $ 4.94 \pm  0.52$ &	 $ 3.05 \pm  0.17$ &	 $ 3.15 \pm  0.14$ &	 $ 3.16 \pm  0.15$ &	 $ 4.21 \pm  0.28$ &	 $ 4.97 \pm  0.24$ \\ 
MB1 &	 $ 4.46 \pm  0.45$ &	 $ 3.09 \pm  0.17$ &	 $ 3.19 \pm  0.16$ &	 $ 5.39 \pm  0.22$ &	 $ 4.11 \pm  0.27$ &	 $ 5.78 \pm  0.22$ \\ 
\enddata
\tablecomments{The mean and standard deviation are taken over
$t/\torb=1.5-2$ for HD models and $t/\torb=3-4$ for MHD models.
See Sections \ref{sec:theory} and \ref{sec:evol} for definition of each quantity.
All velocities are in units of km/s.
Figure~\ref{fig:eratio} for detailed statistical information about energies related to each velocity component.}
\end{deluxetable}

Figures~\ref{fig:evol1} and \ref{fig:evol2} plot time evolution of selected
diagnostics that describe overall disk properties and energetics.  In
Figure~\ref{fig:evol1}, we plot (a) SFR surface density $\SigSFR$, (b)
midplane number density $\rhomid=\overline{\rho}_0/(1.4\mh)$ and (c)
mass-weighted mean height $\abrackets{|z|}_M$.  Figure~\ref{fig:evol2} plots
mass weighted means of (a) three-dimensional turbulent velocity $\vrms{\delta
v}$, (b) sound speed $\vrms{\vth}$, and (c) turbulent $\vrms{\delta {v}_A}$ and
(d) mean $\vrms{\overline{v}_A}$ Alfv\'en velocities.  Here, we subtract the
azimuthal velocity arising from the background shear (not the horizontal
average) for turbulent velocity, $\delta \mathbf{v}\equiv \mathbf{v}+q\Omega
x\yhat$.  The mean and standard deviations of values from
Figure~\ref{fig:evol1} are listed in  Table~\ref{tbl:mean}.  Time averages are
taken over the ``saturated state" time interval of $(t_1, t_2)$.  We also list
in Table~\ref{tbl:mean} the mean and standard deviation of $H=\Sigma/2\rho_0$,
$\zeta_d=\abrackets{|z|}_M/2H$, and $\chi=4\zeta_d\rhosd/\rho_0$.  In
Table~\ref{tbl:meanv}, we report the mean and standard deviations of the
velocities shown in Figure \ref{fig:evol2}, as well as
$\sigma_z=[\vrms[]{\vth}+\vrms[]{v_z}]^{1/2}$, and
$\sigmaeff=[\sigma_z^2+\vrms[]{v_A}/2-\vrms[]{v_{A,z}}]^{1/2}$.

Comparing Models HS (yellow) and HL (black), Figure~\ref{fig:evol1} shows
distinct differences at early stages ($t<\torb$), but no systematic differences
in mean values of physical quantities at later stages.  In Model HS, initial
vertical collapse at $t\sim20\Myr$ triggers an abrupt increase of $\rhomid$ and
hence $\SigSFR$, which then produces feedback that causes a strong reduction in 
$\rhomid$ and $\SigSFR$, leading to further bounces.  These exaggerated
vertical oscillations are a direct consequence of the lack of turbulence in the
initial conditions and early evolution, before feedback has developed.  In
contrast to Model HS, Model HL (and all magnetized models) shows no strong
oscillation at early stages, although more limited vertical oscillation emerges
and persists at later times.  The early driven turbulence and constant heating
in Model HL (and magnetized models) prevent strong, global vertical
oscillations, keeping the midplane density and the thickness of the disk more
or less constant. Without initial vertical collapse, there is no bursting star
formation; rather, the SFR in Model HL remains moderate.  After one orbit time,
when the turbulence driving and heating are fully self-consistent with feedback
from the star formation, all diagnostics in the unmagnetized models quickly
saturate. The convergence of  Models HS and HL to the same saturated state,
despite their completely different early evolution, confirms the robustness of
our previous work. 

The magnetized models also achieve a quasi-steady state, but not so rapidly as
the unmagnetized models. Even after saturation of thermal and turbulent
velocities (both $\delta v$ and $\delta v_A$) as well as the midplane density
and scale height at $\sim 2 \torb$, clear secular evolution continues for the
mean Alfv\'en velocity (or mean magnetic fields; see
Figure~\ref{fig:evol2}(d)).  The models with initially strongest magnetic
fields (MA1 and MB1) slowly lose energy from the mean magnetic field as buoyant
magnetic fields escape through the vertical boundaries.  Conversely, the mean
magnetic energy of models with initially weakest magnetic fields (MA100 and
MA10) slowly grows.  We note, however, that since our horizontal dimension is
only 512 pc and assumed to be periodic, magnetic energy loss might be somewhat
overestimated in our simulations compared to the case in which azimuthal fields
are anchored at larger scales.  In principle, if numerical reconnection in our
models is faster than realistic small-scale reconnection (which is uncertain)
should be, we might also overestimate the growth of mean magnetic fields in
weak-field models. 
Modulo these potential numerical effects, the interesting tendency seen in
Figure~\ref{fig:evol2}(d) is that $\vrms{\overline{v}_A}$ converges toward 
similar values for cases with widely varying initial magnetic fields.

The magnetized models show distinguishably different final saturated states
compared to the unmagnetized models. The SFR surface density and turbulent and
thermal velocity dispersions are lower in the magnetized models, while
variations among the set of magnetized models is small.  This is completely
consistent with expectations from the equilibrium theory, in which (1) the sum
of all pressures must offset a given ISM weight, so the addition of magnetic
pressure reduces the need for turbulent and thermal pressure, and (2) an
increase in ``feedback yields'' due to magnetic fields implies that a lower
star formation rate is needed for equilibrium.  We examine this issue in detail
in Section~\ref{sec:sfr}.

Similarities and differences among models are clearest in the energetics.  In
Figure~\ref{fig:evol2}, we clearly see the saturation of turbulent and thermal
velocity dispersions for all models immediately after $\torb$.  The turbulent
Alfv\'en velocity also converges rapidly except in Model MA100, which converges
after $3\torb$.  Model MB10 achieves a quasi-steady state earliest among the
magnetized models since it has initial magnetic energy comparable to that of
the final state. We thus consider Model MB10 as the fiducial run.  All other
magnetized models converge toward the same saturated state as Model MB10.  As
seen in Figure~\ref{fig:evol2} and Table~\ref{tbl:meanv}, the saturated-state
values of $\vrms{\delta {v}}$, $\vrms{\vth}$, and $\vrms{\delta {v}_A}$, are
essentially indistinguishable for all magnetized models, whereas
$\vrms{\overline{v}_A}$ values show some variations as they have not yet
reached asymptotic values.

\begin{figure}
\plotone{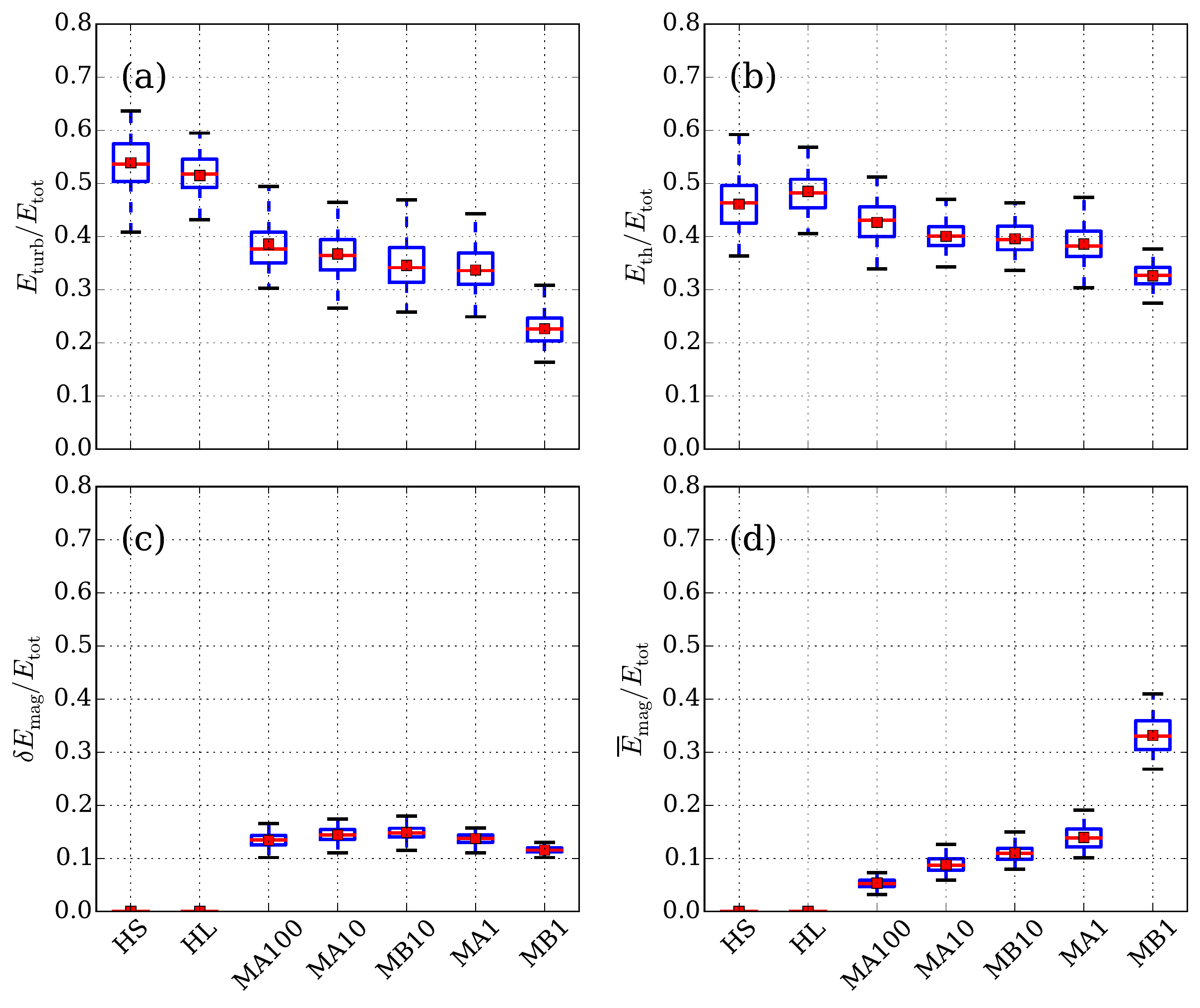}
\caption{Box and whisker plot of the energy ratios (a) turbulent (kinetic)
$\Ekin/E_{\rm tot}$, (b) thermal $\Eth/E_{\rm tot}$, (c) turbulent magnetic
$\Emagt/E_{\rm tot}$, and (d) mean magnetic $\Emago/E_{\rm tot}$, averaged over
$(t_1, t_2)$.  The bottom and top of the rectangular box denote the first
($Q_1$) and third ($Q_3$) quartiles, respectively, while the red line within
the box is the median (second quartile; $Q_2$) of the data.  The lower/upper
whisker reaches to $1.5\times(Q_3-Q_1)$ or the minimum/maximum of the data,
whichever is larger/smaller.  }
\label{fig:eratio}
\end{figure}
 
In  Figure~\ref{fig:eratio}, we summarize the saturated-state energy ratios
with ``box and whisker" plots omitting outliers, and with the mean values (red
squares).  From left to right in each panel, the initial degree of
magnetization increases.  Panels show the fraction of total energy in each
component: turbulent kinetic, thermal, turbulent magnetic, and mean magnetic.
The turbulent kinetic energy is defined by $\Ekin\equiv\sum \Delta
V\rho|\delta\mathbf{v}|^2/2 $, and the thermal energy is $\Eth = \sum \Delta
V\Pth/(\gamma-1) $, where $\gamma=5/3$ in our simulations.  The mean and
turbulent magnetic energies are given by $\Emago\equiv \sum \Delta V
|\overline{\mathbf{B}}|^2/8\pi $ and $\Emagt\equiv \sum \Delta
V|\delta\mathbf{B}|^2/8\pi $.  Since $\overline{\delta \mathbf{B}}=0$ by
definition, $E_{\rm mag}=\Emago+\Emagt$.  Note that these energies are
integrated over the whole volume, which compared to the midplane (see
Section~\ref{sec:equil} and Table~\ref{tbl:meanP}) increases the relative
proportion of thermal to turbulent energy. 

For the unmagnetized models, the turbulent kinetic and thermal energies are
nearly in equipartition.  For Model MB10, which is the most saturated
magnetized model, the total energy is portioned into kinetic (35\%), thermal
(39\%), turbulent magnetic (15\%), and mean magnetic (11\%) terms. Thus,
kinetic, thermal, and magnetic components are each close to 1/3 of the total
energy.  The ratio between turbulent kinetic and turbulent magnetic energies is
about $\Ekin: \Emagt=7:3$, similar to what has been found for saturation of
small scale dynamo simulations at large Reynolds number \citep[e.g.][see
Section~\ref{sec:summary} for details]{2004PhRvE..70a6308H}.  Although the mean
magnetic energy is not yet fully saturated (see Figure~\ref{fig:evol2}(d)) for
all models, there is a clear trend of convergence toward the saturated state of
Model MB10.  Figure~\ref{fig:evol2} and Table~\ref{tbl:meanv} show that
component energies in the magnetized models are nearly the same except the mean
magnetic term, which still reflects initial mean field strengths.
 
It is of interest to characterize the angular momentum transport by both
turbulence and magnetic fields in our models.  We measure the $R$-$\phi$
component of Reynolds and Maxwell stresses as a function of $z$,
$R_{xy}\equiv\overline{\rho v_x \delta v_y}$ and $M_{xy}\equiv\overline{B_x
B_y/4\pi}$, respectively.\footnote{ We confirm that the other off-diagonal
terms of stress tensors are one or two orders of  magnitude smaller.} Since the
stresses are non-negligible only within one gas scale height, it is most
informative to calculate the mass-weighted mean values of the stresses
normalized by the mean midplane thermal pressure, the
``$\alpha_\mathrm{ss}$-parameters'' of \citet{1973A&A....24..337S}:
$\alpha_\mathrm{ss,R}=\abrackets{R_{xy}}_M/\overline{P}_{\rm th,0}$ and
$\alpha_\mathrm{ss,M}=\abrackets{M_{xy}}_M/\overline{P}_{\rm th,0}$ (see
Table~\ref{tbl:meanP} for $\overline{P}_{\rm th,0}$).  As listed in
Table~\ref{tbl:mean}, the $\alpha_\mathrm{ss}$-parameters are comparable to
each other and $\sim 0.2-0.3$.  Although the ratio between Reynolds and Maxwell
stresses is completely different compared to the the case of turbulence driven
by magnetorotational instability (where the Maxwell stress dominates), the
total stress $\sim 0.4-0.5$ is similar
\citep[e.g.,][]{1995ApJ...440..742H,2003ApJ...599.1157K,2005ApJ...629..849P}.
Note that gravitational stress in our simulations is generally lower, with
$\alpha_\mathrm{ss}\sim 0.05$.  This is also true for other simulations in
which turbulence is driven by gravitational instability combined with sheared
rotation, which give $\alpha_\mathrm{ss}$ less than $0.1$
\citep[e.g.,][]{2014ApJ...789...34S}.  The total stress gives
$\alpha_\mathrm{ss}\sim0.4-0.5$, implying that the gas accretion time $t_{\rm
acc}\sim R^2\Omega/(\alpha_{\rm ss} \vth^2)=390\Gyr$ for $\vth=3\kms$ using
$R=8$~kpc.

\begin{figure}
\plotone{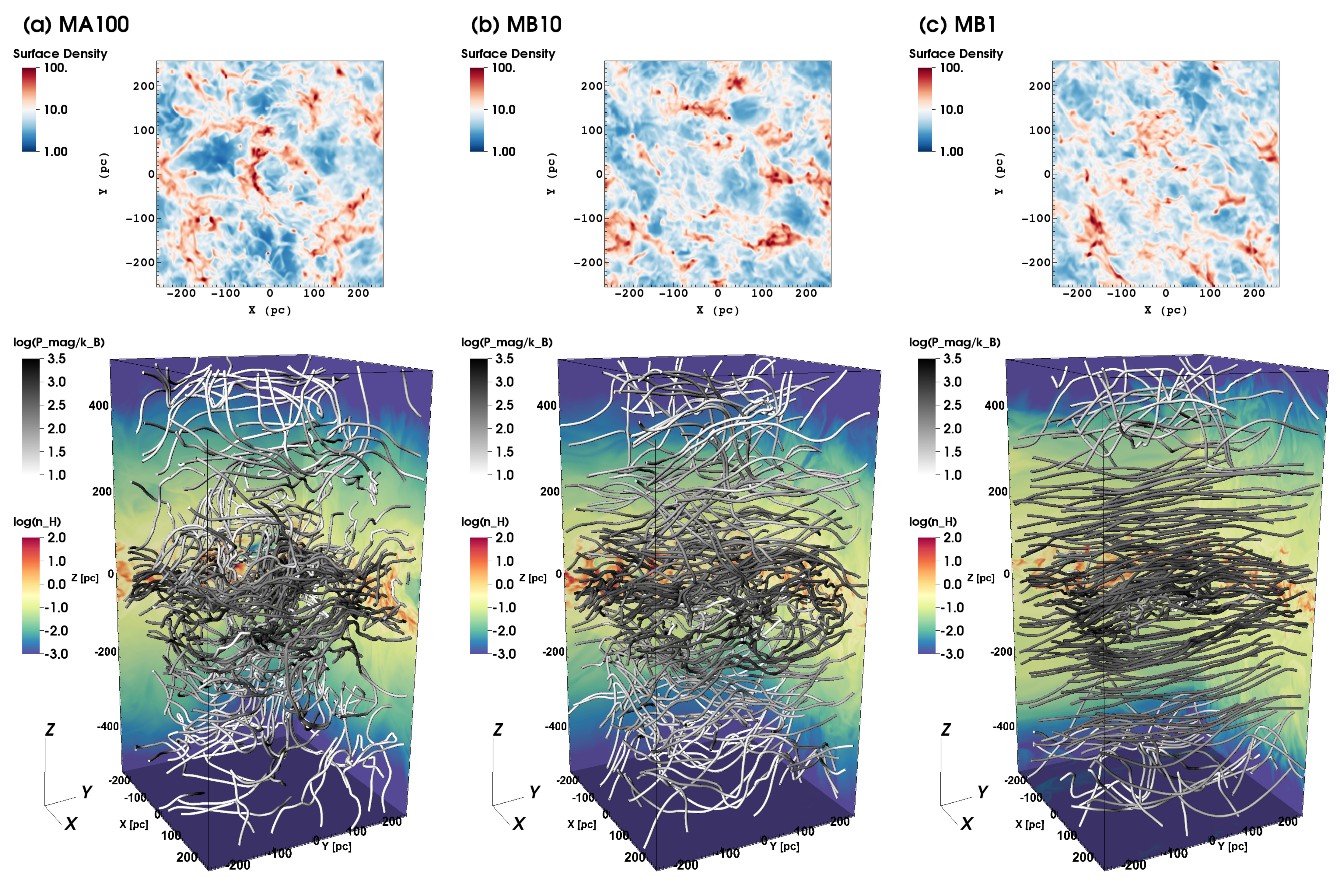}
\caption{Gas and magnetic structure in Models (a) MA100, (b) MB10, and (c) MB1
at $t/\torb=3$.  Top: Surface density in units of $\Surf$. Bottom: 3D
visualization of magnetic field lines, with vertical slices of hydrogen number
density shown in color scale on box boundaries.  Model MB1 shows field lines
preferentially along $\yhat$ within $|z|<300\pc$ since its mean magnetic fields
are still dominant. From right to left, randomness of the magnetic field
structure increases as the mean field strength decreases, since the turbulent
field strengths are all similar at this time.  Differences in the magnetic
field produce no clear signature in surface density maps, however, with similar
cloudy structure in all cases.  }
\label{fig:sat}
\end{figure}

To illustrate the overall saturated-state disk structure, in
Figure~\ref{fig:sat} we display surface density (top) and  magnetic field lines
(bottom) for Models (a) MA100, (b) MB10, and (c) MB1 at $t/\torb=3$.  As shown
in Figures~\ref{fig:evol2} and \ref{fig:eratio}, the strength of the mean
magnetic field increases from MA100 to MB10 to MB1 (from left to right in
Figure~\ref{fig:sat}).  Averaged over the saturation period ($t/\torb=3-4$),
the mean and turbulent magnetic field values at the midplane are
$\overline{\mathbf{B}}=(-0.26,1.2,0)\muG$
and $\delta\mathbf{B}_{\rm rms}=(1.3,1.6,1.1)\muG$ for Model MA100,
$\overline{\mathbf{B}}=(-0.30,1.7,0)\muG$
and $\delta\mathbf{B}_{\rm rms}=(1.4,1.7,1.2)\muG$ for Model MB10, and
$\overline{\mathbf{B}}=(-0.15,2.5,0)\muG$
and $\delta\mathbf{B}_{\rm rms}=(1.4,1.8,1.2)\muG$ for Model MB1.
For all models, the azimuthal ($\hat y$) component is the largest of
$\overline{\mathbf{B}}$.  However, this component exceeds the turbulent
components only for model MB1; for model MB10 it is comparable to the largest
turbulent component, and for model MB100 it is smaller than the largest
turbulent component.  As a result, field lines are more complex and random in
Model MA100 ($\Emagt>\Emago$) and more aligned in a preferential direction
(along $\yhat$) in Model MB1 ($\Emagt<\Emago$).  The dominance of the mean
magnetic fields at all heights in Model MB1 is also evident.

Despite of the strong distinctions in the structure of field lines, the surface
density maps look quite similar for all models.  In particular, there is no
visually prominent evidence of alignment of dense filaments either
perpendicular or parallel to the mean magnetic field direction.  However,
traces of the shear are evident in overall pattern of striations (consistent
with trailing wavelets), particularly for Model MB1.  More quantitative
analysis of the  morphology of filaments and magnetic field lines may be
obtained using maps of synthetic 21 cm emission, dust emission, and
polarization \citep[e.g.,][]{2013ApJ...774..128S,2015arXiv150204123P}.  We
defer this interesting study to future work.

\subsection{Vertical Dynamical Equilibrium}\label{sec:equil}

In this subsection, we investigate the vertical dynamical equilibrium of model
disks using horizontally and temporally averaged profiles of $\Pturb$, $\Pth$,
$\Pmagt$, and $\Pmago$, in comparison to profiles of the ISM weight
$\mathcal{W}$.  We also compare midplane values.  

\begin{figure}
\plotone{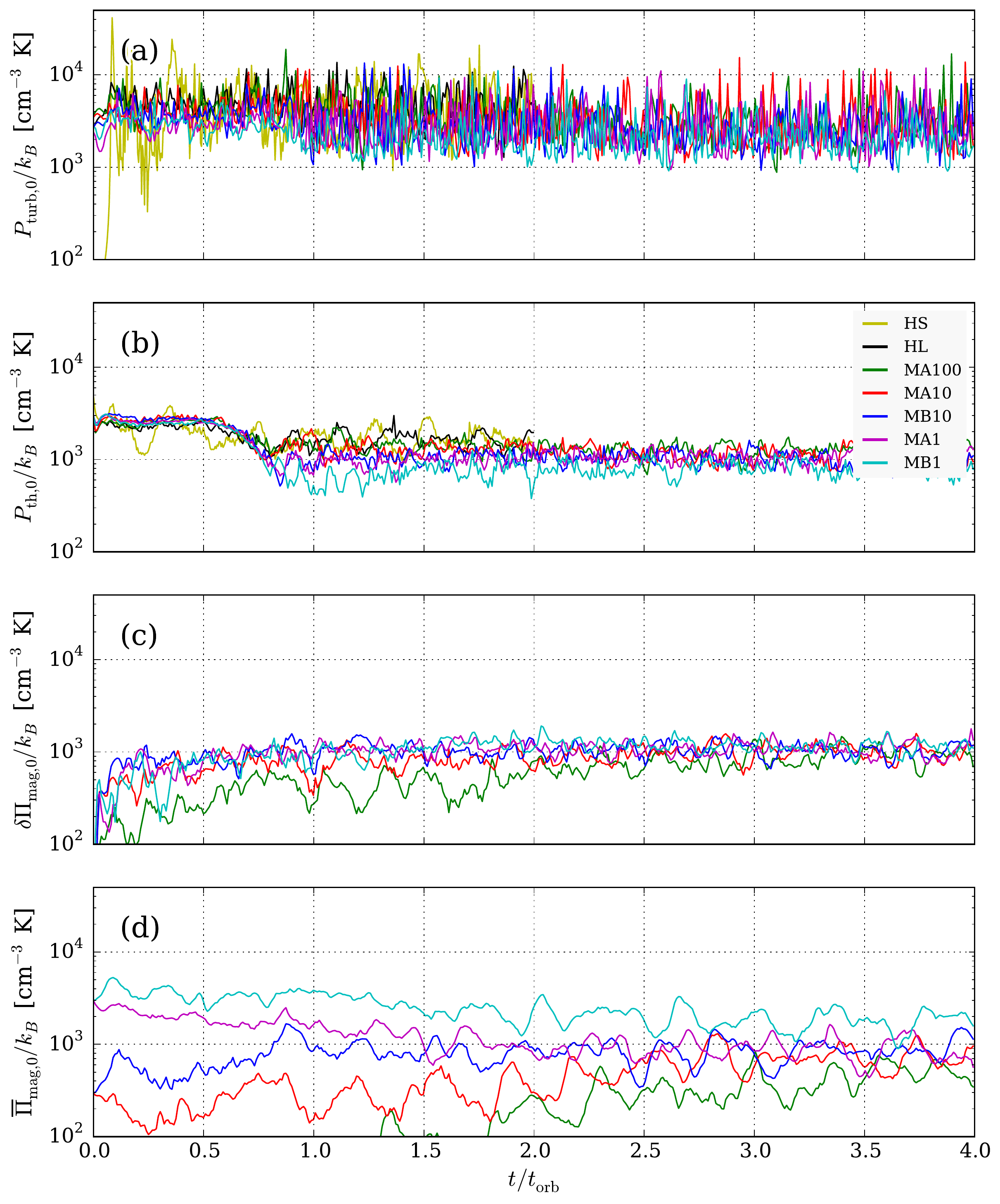}
\caption{Time evolution of horizontally-averaged components of support at the
midplane: (a) turbulent kinetic $\Pturb$, (b) thermal $\Pth$, (c) turbulent
magnetic $\Pmagt$, and (d) mean magnetic $\Pmago$.  }\label{fig:Pevol}
\end{figure}

\begin{deluxetable}{lccccccc}
\tablewidth{0pt}
\tabletypesize{\footnotesize}
\tablecaption{Vertical Support and Feedback Yield at Saturation\label{tbl:meanP}}
\tablehead{
\colhead{Model} &
\colhead{$\Pturb$} &
\colhead{$\Pth$} &
\colhead{$\Pmagt$} &
\colhead{$\Pmago$} &
\colhead{$\etaturb$} &
\colhead{$\etath$} &
\colhead{$\etamagt$}} 
\startdata
HS &	 $ 5.24 \pm  3.05$ &	 $ 1.78 \pm  0.40$ &	 \nodata           &	 \nodata           &	 $ 3.30 \pm  2.08$ &	 $ 1.12 \pm  0.37$ &	 \nodata           \\
HL &	 $ 5.15 \pm  2.61$ &	 $ 1.61 \pm  0.29$ &	 \nodata           &	 \nodata           &	 $ 3.53 \pm  1.91$ &	 $ 1.10 \pm  0.29$ &	 \nodata           \\
MA100 &	 $ 3.40 \pm  2.42$ &	 $ 1.24 \pm  0.18$ &	 $ 0.94 \pm  0.21$ &	 $ 0.46 \pm  0.15$ &	 $ 3.72 \pm  2.73$ &	 $ 1.35 \pm  0.31$ &	 $ 1.03 \pm  0.29$ \\
MA10 &   $ 3.38 \pm  2.66$ &     $ 1.14 \pm  0.20$ &     $ 1.04 \pm  0.18$ &     $ 0.73 \pm  0.18$ &     $ 4.17 \pm  3.38$ &     $ 1.41 \pm  0.36$ &     $ 1.29 \pm  0.33$ \\
MB10 &	 $ 2.85 \pm  1.46$ &	 $ 1.04 \pm  0.18$ &	 $ 1.03 \pm  0.17$ &	 $ 0.87 \pm  0.23$ &	 $ 3.83 \pm  2.12$ &	 $ 1.40 \pm  0.38$ &	 $ 1.38 \pm  0.37$ \\
MA1 &    $ 3.02 \pm  2.11$ &     $ 1.05 \pm  0.16$ &     $ 1.10 \pm  0.22$ &     $ 0.91 \pm  0.29$ &     $ 4.00 \pm  2.92$ &     $ 1.39 \pm  0.37$ &     $ 1.46 \pm  0.43$ \\
MB1 &	 $ 2.18 \pm  1.12$ &	 $ 0.83 \pm  0.16$ &	 $ 1.16 \pm  0.18$ &	 $ 1.83 \pm  0.45$ &	 $ 3.92 \pm  2.23$ &	 $ 1.49 \pm  0.47$ &	 $ 2.08 \pm  0.61$ \\
\enddata
\tablecomments{The mean and standard deviation are taken over 
$t/\torb=1.5-2$ for HD models and $t/\torb=3-4$ for MHD models.
The vertical support terms at the midplane ($\Pturb$, $\Pth$, $\Pmagt$, and $\Pmago$) 
are given in units of $10^3\kbol\Punit$.
See Equation~(\ref{eq:eta}) for $\eta$ definition and 
Figure~\ref{fig:eta} for detailed statistical information about feedback yields.}
\end{deluxetable}

Figure~\ref{fig:Pevol} plots time evolution of horizontally-averaged midplane
support terms: (a) $\Pturb$,  (b) $\Pth$,  (c) $\Pmagt$, and (d) $\Pmago$. The
mean and standard deviation values over $(t_1, t_2)$ are summarized in
Table~\ref{tbl:meanP}. Table~\ref{tbl:meanP} also lists the feedback yields for
each support component in the units of Equation (\ref{eq:eta}).  Similar to
Figures~\ref{fig:evol1} and \ref{fig:evol2}, convergence of each support term
except the mean magnetic field is evident after $t/\torb>3$, regardless of
initial magnetic field strength.  The mean magnetic support $\Pmago$ in
Figure~\ref{fig:Pevol}(d) more gradually converges toward the value of the
fiducial run, Model MB10.

\begin{figure}
\plotone{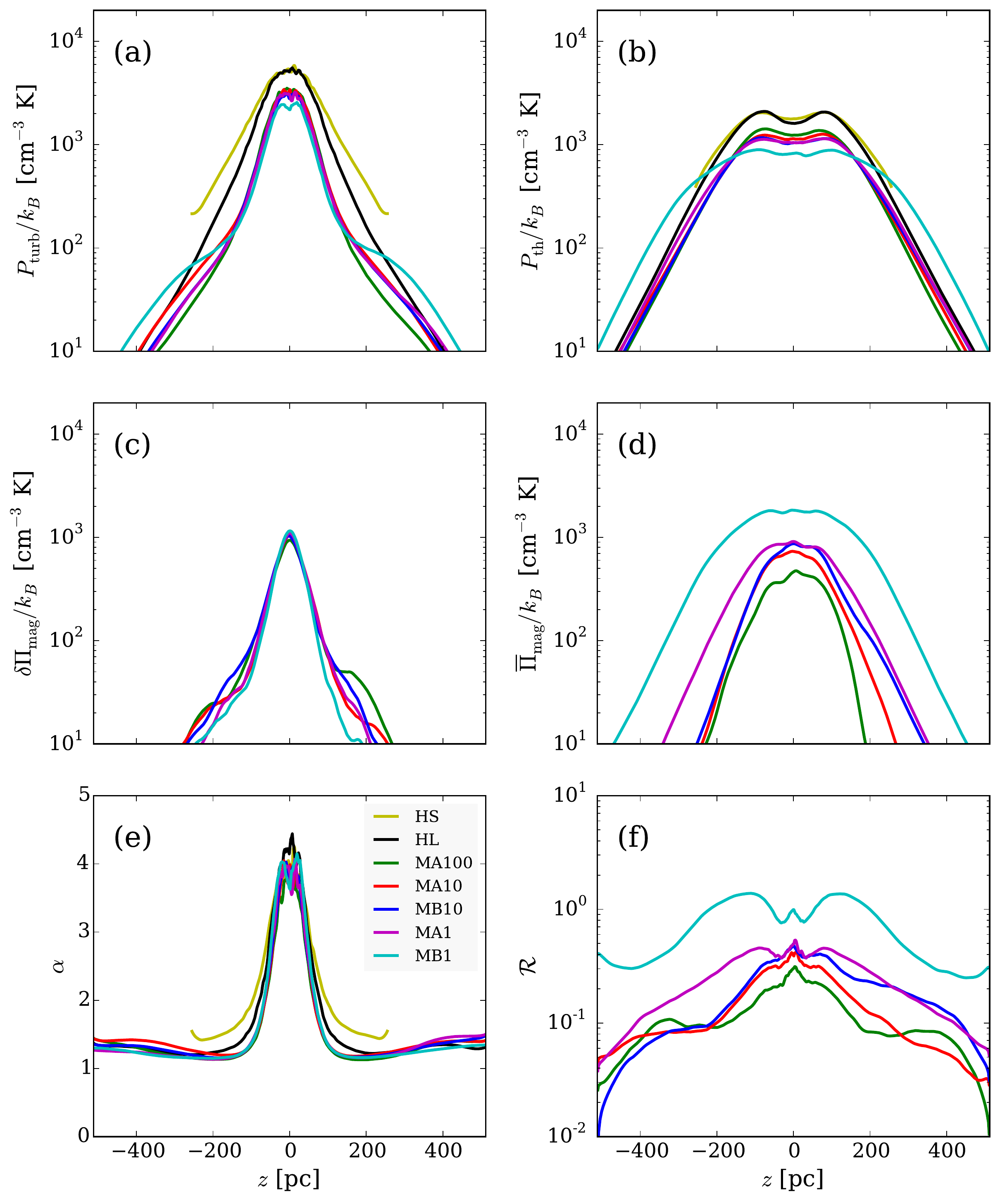}
\caption{ Temporally and horizontally averaged vertical profiles of (a)
$\Pturb$, (b) $\Pth$, (c) $\Pmagt$, (d) $\Pmago$, (e)
$\alpha\equiv(\Pturb+\Pth)/\Pth$, and (f) $\mathcal{R}\equiv
(\Pmagt+\Pmago)/(\Pturb+\Pth)$.
\label{fig:Pprof}}
\end{figure}

Figure~\ref{fig:Pprof}(a)-(d) plots vertical profiles of the four individual
support terms averaged over $(t_1, t_2)$ and the horizontal direction.  Panel
(e) shows the ratio of the kinetic (turbulent + thermal) to the thermal term,
$\alpha\equiv(\Pturb+\Pth)/\Pth$; and panel (f) shows the ratio of total
magnetic to kinetic term, $\mathcal{R}\equiv (\Pmagt+\Pmago)/(\Pturb+\Pth)$. 

First of all, Figure~\ref{fig:Pprof} confirms that Models HS and HL agree very
well not only for the mean and midplane values, but also for the overall
profiles. The periodic vertical boundary conditions in Model HS introduce a
small anomaly near the vertical boundaries.  

In the presence of magnetic support, the turbulent and thermal terms are both
reduced. However, the relative contribution between turbulent and thermal terms
remains similar (Figure~\ref{fig:Pprof}(e)): $\alpha\sim 4$ within one scale
height, decreasing to $\alpha\sim 1$ at high-$|z|$. This is an important
consequence of self-regulation by star formation feedback, explained in the
equilibrium theory (see Section~\ref{sec:sfr}). The turbulent magnetic support
in all magnetized models converges to very similar profiles
(Figure~\ref{fig:Pprof}(c)), while the mean magnetic support still shows
differences (Figure~\ref{fig:Pprof}(d)), especially for two extreme cases
(Models MA100 and MB1).  For all models except MB1, Figure~\ref{fig:Pprof}(f)
shows that the midplane value $\mathcal{R}_0\sim$0.2-0.5, while $\mathcal{R}$
becomes very small at high-$|z|$; Model MB1 has $\mathcal{R}\sim 0.3-1$. 

\begin{figure}
\plotone{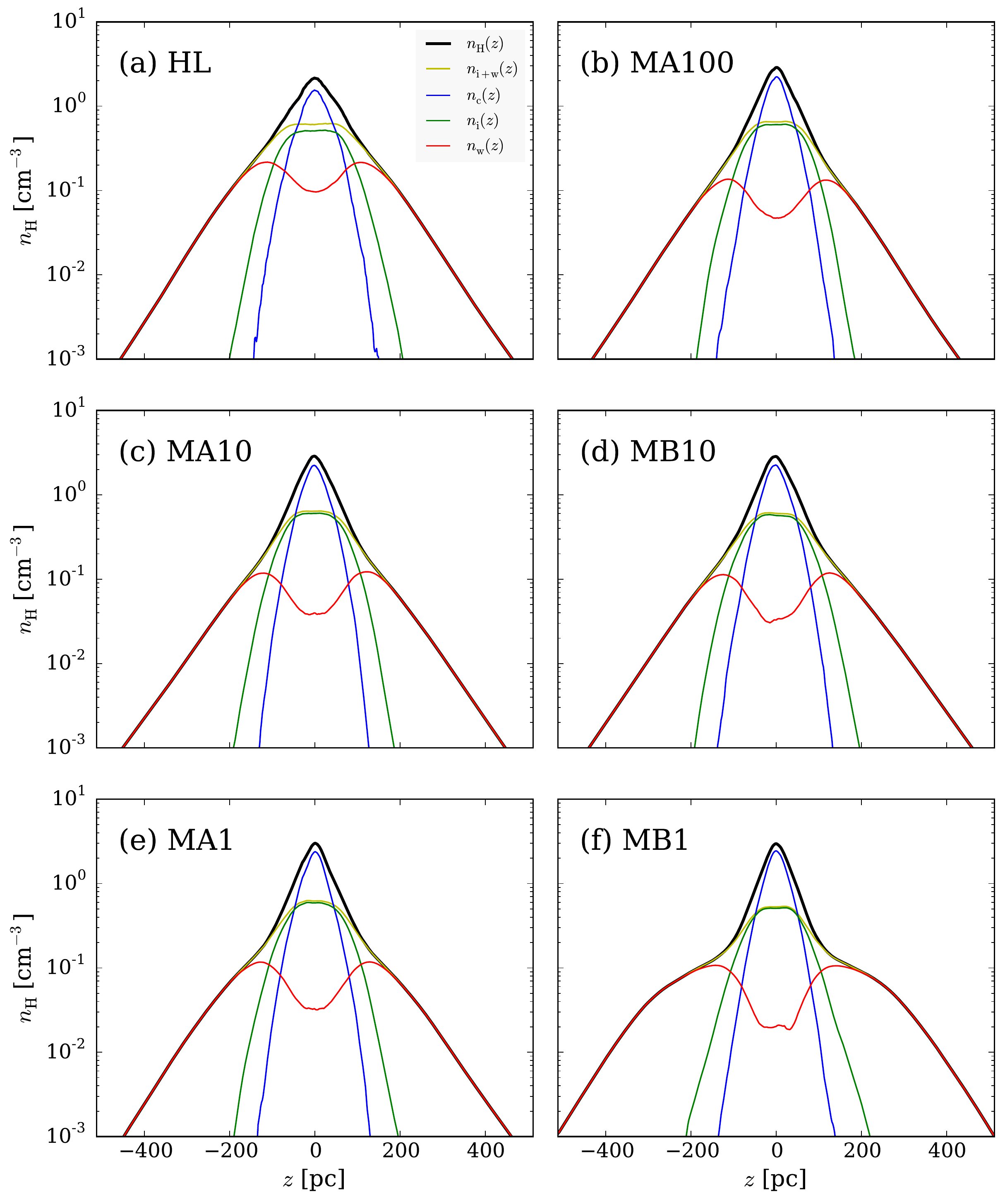}
\caption{
Vertical saturated-state density profiles of cold (blue; $T<184\Kel$),
intermediate-temperature (green; $184\Kel<T<5050\Kel$), and warm (red;
$T>5050\Kel$) gas phases.  The profiles of total density (black) and combined
non-cold phases (yellow; $T>184\Kel$) are also shown.  \label{fig:density}}
\end{figure}

Figure~\ref{fig:density} presents vertical profiles of the
horizontally-averaged density from different thermal components of gas, for
each model.  We separately plot the profiles of cold ($n_c$; $T<184\Kel$),
intermediate-temperature ($n_i$; $184\Kel<T<5050\Kel$), and warm ($n_w$;
$T>5050\Kel$) phases, as well as the whole medium ($n_H=n_c+n_i+n_w$) and
combined non-cold ($n_{i+w}$; $T>184\Kel$) components.  The distribution of
cold medium is mostly limited to one gas scale height, while the warm medium
extends to high $|z|$ with a dip near the midplane.  The intermediate
temperature gas is not as concentrated toward the midplane as the cold medium,
but does not extend to high $|z|$.  Decomposition into cold ($n_c$) and
non-cold ($n_{i+w}$) components gives two smooth profiles that resemble the
results for two-component fits to observed \ion{H}{1} gas from 21 cm emission.

Because most of cold medium resides within one gas scale height (and we do not
consider runaway O stars), most of SN explosions occur there.  The $\Pturb$ and
$\alpha$ profiles in Figure~\ref{fig:Pprof} are peaked near the midplane, with
scale heights similar to the cold medium scale height.  The value of $\alpha$
is close to unity at $|z| > 100$pc, implying that much of the turbulent energy
dissipates very efficiently within the driving layer without propagating to
high $|z|$.  Thermal pressure is flat in the central layer, implying that the
two-phase medium is well-mixed and in pressure equilibrium.  Beyond the
turbulent, two-phase layer, the ISM is mostly warm gas and is supported mainly
by thermal pressure.  Note that the shape of vertical profiles of the Reynolds
and Maxwell stress are respectively similar to those of $\Pturb$ and $\Pmagt$
(see Figure~\ref{fig:Pprof}(a) and (c)).

\begin{figure}
\plotone{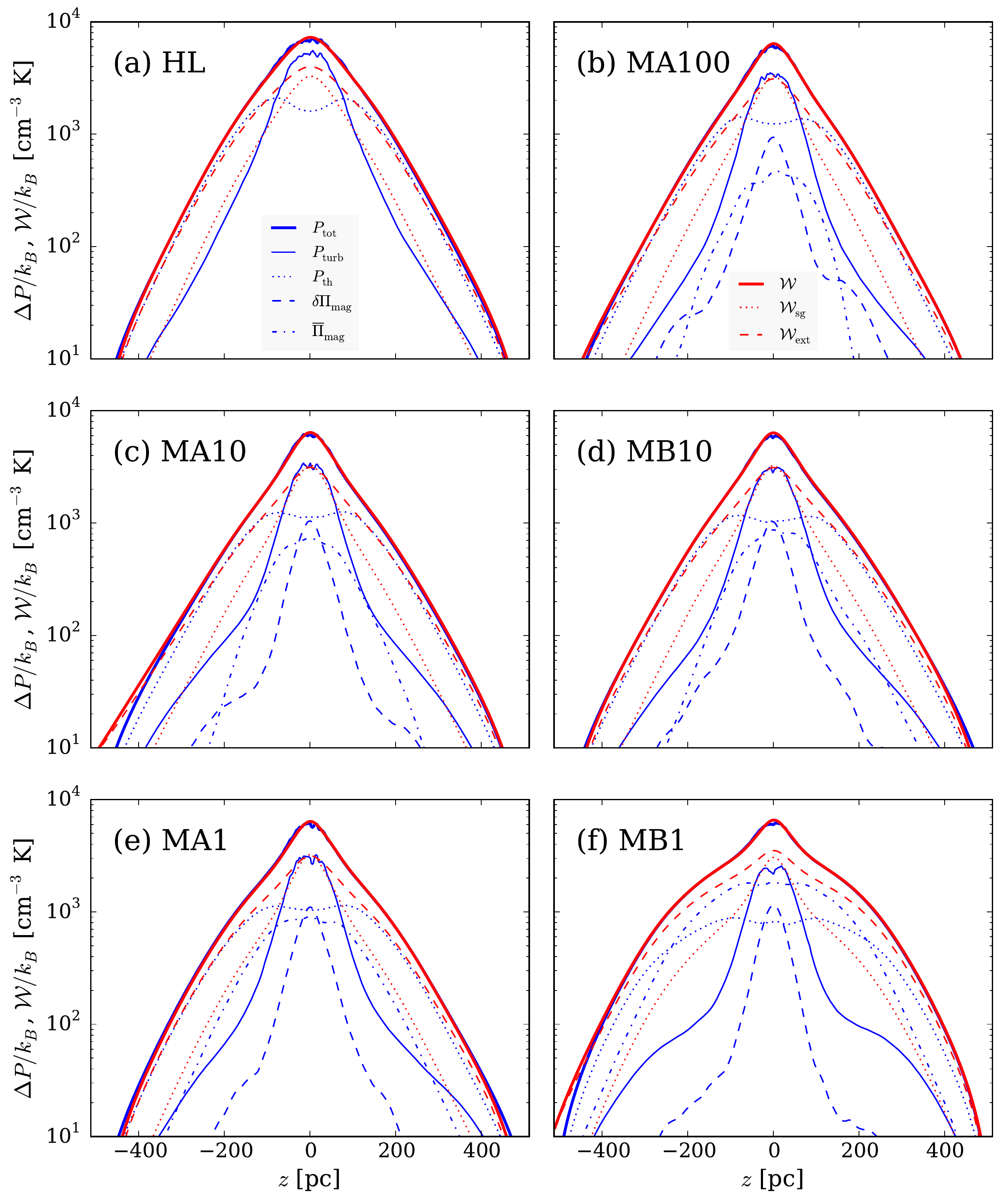}
\caption{ Profiles of saturated-state vertical support (blue) and gas weight
(red). Total support and total weight are shown in thick lines.  Thin blue
solid, dotted, dashed, dot-dashed lines denote turbulent kinetic, thermal,
turbulent magnetic, and mean magnetic support, respectively.  Thin red dotted
and dashed lines denote the weight of the ISM under self- and external gravity,
respectively.  \label{fig:balance}}
\end{figure}

In order to check vertical dynamical equilibrium quantitatively, for each model
we calculate the weight of the gas using the horizontally and temporally
averaged density and gravitational potentials, with $\Wsg(z)$ and $\Wext(z)$ as
in Equations (\ref{eq:wsg}) and (\ref{eq:wext}) except integrated between $z$
and $\zmax$.  Figure~\ref{fig:balance} plots the vertical profiles for the
total support $\Delta\Ptot(z) = {\Ptot}(z) - {\Ptot}(\zmax)$ (blue) and the
weight $\mathcal{W}(z)=\Wsg(z)+\Wext(z)$ of the ISM (red).  We also plot each
component of the vertical support shown in Figure~\ref{fig:balance} along with
the individual weights, to show the relative importance of the contributing
terms in each model.

Figure~\ref{fig:balance} shows that vertical equilibrium is remarkably well
satisfied (thick blue and red lines overlap almost completely).  In the present
simulations, the magnetic scale height is not very different from the gas scale
height, especially for the turbulent component.  Thus, similarly to the gas
pressure terms we can simply replace $\Delta\Pmag\rightarrow {\Pmag}_{,0}$ in
Equations (\ref{eq:Ptot}) and (\ref{eq:R}).\footnote{Note, however, that in
observations the mean magnetic field has a scale height much larger than that
of the warm/cold ISM.  This may be due to effects (not included in the present
models) that help drive magnetic flux out of galaxies, including a hot ISM and
cosmic rays.  For this reason, Equations (\ref{eq:Ptot}) and (\ref{eq:R}) are
most generally written in terms of differences in the magnetic support, and
also allow for differences in radiation and cosmic ray pressure across the
warm/cold ISM layer \citep{2011ApJ...731...41O}.} Since the weight of the gas
(RHS of Eq. (\ref{eq:de})) is more or less the same for all of the present
models ($\chi\sim1$ within 25\% from Table~\ref{tbl:mean}), the additional
support from both the mean and turbulent magnetic fields necessarily implies a
reduction in the turbulent and thermal pressures compared to unmagnetized
models.  Although external gravity dominates the weight at high-$|z|$, self-
and external gravity contributions are almost the same close to the midplane.
Within the turbulent driving layer, turbulent pressure dominates other support
terms for all models, while the thermal pressure dominates at high-$|z|$.  The
turbulent and mean magnetic support terms are as important as the thermal
pressure at the midplane.  Only for Model MB1, the mean magnetic support is
substantial at all $z$.  Table~\ref{tbl:meanP} includes the midplane values of
the contribution to vertical support from each component.

\subsection{Regulation of Star Formation Rates}\label{sec:sfr}

\begin{figure}
\plotone{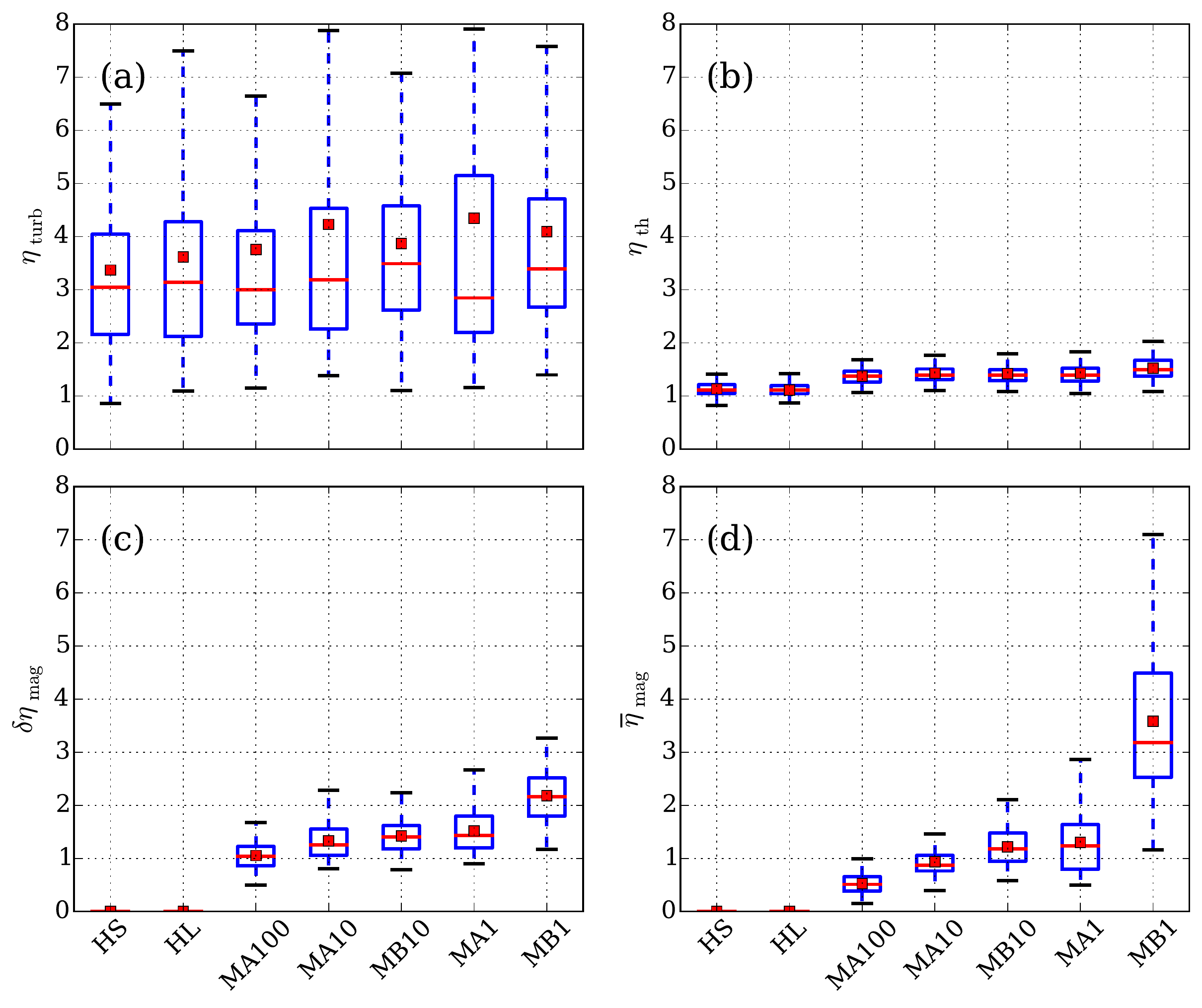}
\caption{Feedback yields, defined in Equation~(\ref{eq:eta}): (a) $\etaturb$,
(b) $\etath$, (c) $\etamagt$, and (d) $\etamago$. Box and whisker plot is as in
Figure~\ref{fig:eratio}.  Note that the mean (square symbol) and median (red
line) of $\etaturb$ are different since the spikes in time evolution of
$\Pturb$ (see Figure~\ref{fig:Pevol}(a)) caused by SN feedback affect the mean
more than the median.  } \label{fig:eta}
\end{figure}

We compute the feedback yield (ratio of midplane support to surface density of
star formation rate) for each component at saturation, and report values in
Table~\ref{tbl:meanP} in the units of Equation~(\ref{eq:eta}).  These yields
are a quantitative measure of the efficiency of feedback for controlling star
formation.  Figure~\ref{fig:eta} presents a box and whisker plot for each
component feedback yield.  As shown in Figure~\ref{fig:eta}, the individual
feedback yields except $\etamago$ are quite similar for all models. Notably,
the magnetized models and unmagnetized models have comparable value for both
turbulent and thermal feedback yields, $\etaturb\sim 3.5-4$ and
$\etath\sim1.1-1.4$.  The agreement of $\etaturb$ between HD and MHD models
stems from the similarity in dissipation timescales for HD and MHD turbulence
\citep[e.g.,][]{1998ApJ...508L..99S,1998PhRvL..80.2754M}.

Except for models MA100 and MB1, the magnetic field and especially the
turbulent magnetic energy become fairly well saturated, due to  efficient
generation from the small-scale turbulent dynamo.  The resulting turbulent
magnetic feedback yield for these models is $\etamagt\sim1.3-1.5$, providing a
ratio $\etaturb/\etamagt\sim 3-4$.  This is equivalent to the energy ratios of
$\Ekin/\Emagt\sim (1.5-2)$ found in driven turbulence MHD simulations
\citep[e.g,][]{1998ApJ...508L..99S,2004PhRvE..70a6308H,2009ApJ...693.1449C,2009ApJ...691.1092L}.
Table~\ref{tbl:meanP} includes $\etamago$ for reference.  However, it should be
kept in mind that the mean magnetic field support is less directly related to
the SFR than the other terms.  Mean magnetic fields arise from the mean-field
dynamo, which depends on the turbulent magnetic field and therefore indirectly
on the SFR, but also depends on other physical effects in a complex manner that
is not well understood.  From the present simulations, we simply note that
$\etamago$ is comparable to $\etamagt$ for the saturated models. 

The theoretical idea of near-linear relationships between SFRs and turbulent
and thermal pressures for self-regulated equilibrium ISM states was introduced
and initially quantified by
\citet{2010ApJ...721..975O,2011ApJ...731...41O,2011ApJ...743...25K}.  In the
present simulations, our momentum feedback prescription for SNe injects radial
momentum of 
$p_* = 3\times 10^5 \Msun\kms$
 within a $10\pc$ sphere (see
Paper~I for details); other work confirms this value for the momentum from a
single SN of energy $10^{51} \erg$, insensitive to the mean value of the
ambient density or cloudy structure in the ambient ISM  \citep[see][and
references within]{2015ApJ...802...99K}.  Our adopted value for the total mass
of new stars formed per SN is $m_*=100\Msun$
\citep[e.g.][]{2001MNRAS.322..231K}. For these feedback parameters, the
momentum flux/area in the vertical direction is then
$\Pdriv=(p_*/4m_*)\SigSFR=3.6\times10^3\SigSFRnorm\kbol\Punit$
\citep{2011ApJ...731...41O};
 this is the
effective turbulent driving rate in the vertical direction.  If dissipation of
turbulence occurs in approximately a vertical crossing time over $H$ (the main
energy-containing scale), the expected saturation level for turbulent pressure
is ${\Pturb}_{,0}\approx\Pdriv$, giving $\etaturb=3.6$.  Direct calibration in
Paper~I gives ${\Pturb}_{,0}/\Pdriv=1.20\SigSFRnorm^{-0.11}$ \citep[see
also][]{2011ApJ...743...25K}.  Including results from both previous and current
simulations, we obtain ${\Pturb}_{,0}/\Pdriv=0.9$-$1.1$ from all HD and MHD
models.  Thus, the turbulent pressure is consistent with theoretical
predictions, insensitive to magnetization.  

As the SFR varies in our simulations, the time-varying heating rates move the
thermal equilibrium curves up and down in the density-pressure phase plane. To
maintain a two-phase medium within the midplane layer, the actual thermal
pressure of gas must change to be self-consistent with the changing heating
rate. Specifically, for a two-phase medium the range of the midplane thermal
pressure is between the minimum and maximum pressures of the cold and warm
medium, $\Pmin$ and $\Pmax$, respectively.  Our adopted cooling and heating
formalism gives ${\Pmin}_{,3}=0.7\SigSFRnorm$ and ${\Pmax}_{,3}=2.2\SigSFRnorm$
\citep[see][]{2002ApJ...564L..97K,2008ApJ...681.1148K}.  The geometric mean of
these two pressures is representative of  the expected thermal pressure in a
two-phase medium \citep[e.g.,][]{1995ApJ...443..152W,2003ApJ...587..278W},
yielding ${\Ptwo}_{,3}\equiv({\Pmin}_{,3}{\Pmax}_{,3})^{1/2}=1.2\SigSFRnorm$.
\citet{2010ApJ...721..975O}
 argued that the self-consistent expected midplane pressure for a
star-forming disk in equilibrium is ${\Pth}_{,0}\sim\Ptwo$, corresponding to
$\etath=1.2$ for the thermal feedback yield.  Here, we obtain
${\Pth}_{,0}/\Ptwo=1.0$-$1.3$ for all HD and MHD models, consistent with the
theory in \citet{2010ApJ...721..975O},
and with the numerical results in Paper~I \citep[see
also][]{2011ApJ...743...25K}, ${\Pth}_{,0}/\Ptwo=1.09\SigSFRnorm^{-0.14}$.

In addition to the turbulent and thermal pressures, the turbulent magnetic
support is also directly related to the SFR.  As demonstrated in Figures
\ref{fig:evol2} and \ref{fig:Pevol}, the small-scale turbulent dynamo generates
turbulent magnetic fields efficiently.  The turbulent magnetic energy is expect
to saturate at roughly equipartition level with turbulent kinetic energy.  In
our simulations, the saturation level of turbulent magnetic energy is $\sim
40\%$ or slightly smaller compared to the turbulent kinetic energy (see
Figure~\ref{fig:eratio}).  If turbulent kinetic and magnetic components are all
isotropic so that $\Pturb=(1/3)\rho |\delta \mathbf{v}|^2=(2/3)\Ekin$ and
$\Pmagt=(1/3)|\delta \mathbf{B}|^2/8\pi=(1/3)\Emagt$, then $\Ekin\sim 2\Emagt$
(Figure \ref{fig:eratio}) would give $\Pturb\sim 4\Pmagt$.  We find
${\Pturb}_{,0}\sim \Pdriv \sim 3 {\Pmagt}_{,0}$, except for Model MB1, see
Table \ref{tbl:meanP}.  In idealized, driven MHD turbulence simulations,
$\Ekin/\Emagt\sim 1-2.5$ for various initial magnetic fields, Mach number, and
compressibility of gas
\citep[e.g.,][]{2004PhRvE..70a6308H,2009ApJ...693.1449C,2009ApJ...691.1092L}.
Our simulations are generally consistent with saturation energy levels from
idealized driven turbulence experiments in a periodic box, keeping in mind that
identical results are not expected considering differences in the setup.  (That
is, rather than an idealized periodic box with spectral driving of turbulence,
our simulations incorporate complex physics to model realistic galactic disks
including vertical stratification, compressibility, spatially localized
turbulent driving, self-consistent evolution of mean fields, self-gravity,
presence of cooling and heating, and so on.)

Our simulations are suggestive, but not conclusive, with respect to the
asymptotic equilibrium state of the mean magnetic support and its connection to
SFRs.  The mean fields for all models appear to be converging to the same level
of support as the turbulent fields (e.g., as in Model MB10; see Figure
\ref{fig:evol1}), which as shown above have ${\Pmagt}\propto\SigSFR$.  Note
that the mean magnetic field is anisotropic and dominated by the azimuthal
component, so $\Pmago = \Emago$ in contrast to $\Pmagt = (1/3) \Emagt$ for the
isotropic case.  However, longer-term simulations would be needed to confirm
convergence of $\Pmago$, because the evolution time scale for the mean field is
much longer than for the turbulent field (see Figure~\ref{fig:evol2}).  In
addition, the mean field level may in principle be affected by numerical
parameters including the horizontal box size and effective numerical
resistivity.  Thus, it remains uncertain whether the mean magnetic support
should be considered as directly related to the SFR or not.  Models MA10, MB10,
and MA1 appear closest to saturation in their mean magnetic field, and have
${\Pmago}_{,0}/\Pdriv\sim 0.3$, which would correspond to $\etamago\sim1$.  For
reference, Models MA100 and MB1 respectively have ${\Pmago}_{,0}/\Pdriv\sim
0.14$ and 0.9, although with the caveat that these models are not
asymptotically converged.

\begin{figure}
\plotone{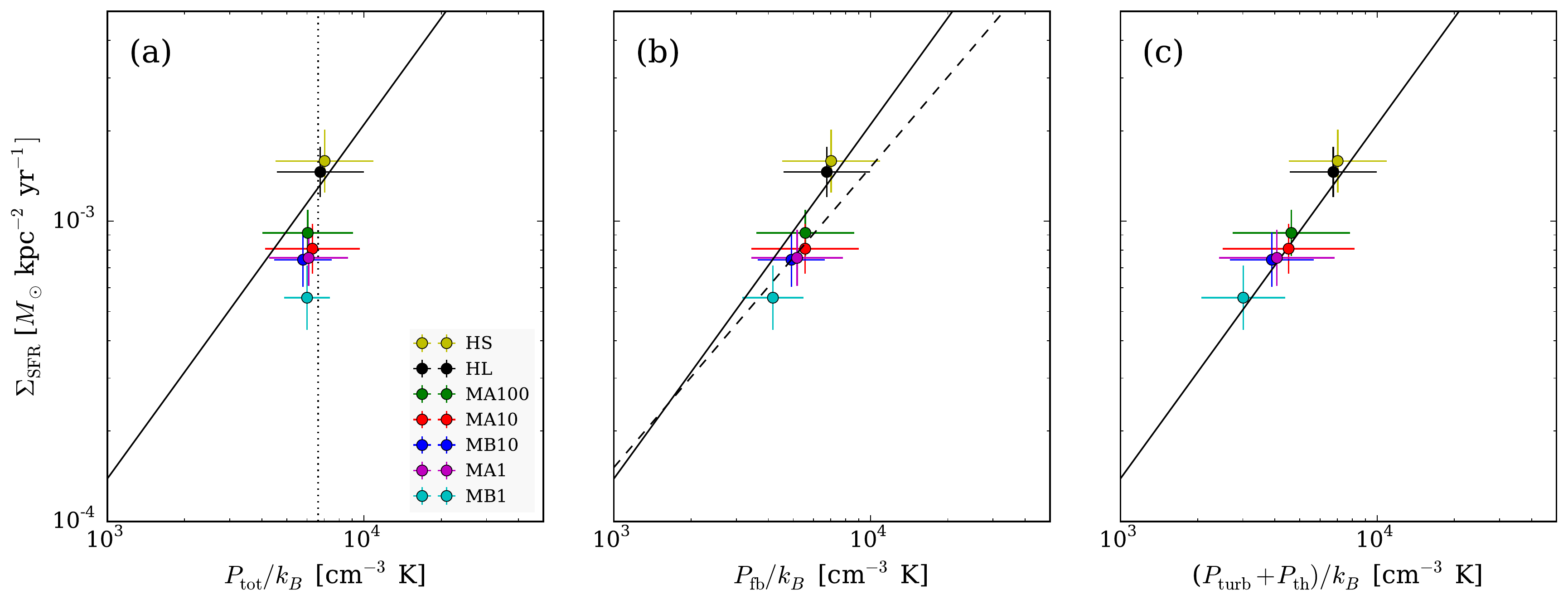}
\caption{The SFR surface density as a function of midplane values of the
effective pressure, (a) total pressure $\Ptot\equiv\Pturb+\Pth+\Pmagt+\Pmago$,
(b) feedback pressure $\Pfb\equiv\Pturb+\Pth+\Pmagt$, and (c) kinetic pressure
$\Pturb+\Pth$.  The solid line denotes the fit from a wide parameter space of
HD models in Paper~I (see text). The vertical dotted line in (a) is the
midplane pressure required for the vertical dynamical equilibrium, $\PDE=\pi G
\Sigma^2(1+\chi)/2=6.6\times10^3\kbol\Punit$ for $\chi=1$.  The dashed line in
(b) is the relationship between $\SigSFR$ and $\Pfb$ using
Equation~(\ref{eq:eta}) with total feedback yield for the fiducial run, MB10,
$\etaturb+\etath+\etamagt=6.6$.  \label{fig:sfr}}
\end{figure}

Figure~\ref{fig:sfr} plots $\SigSFR$ as functions of three midplane (effective)
pressures, (a) total pressure
${\Ptot}\equiv{\Pturb}_{}+{\Pth}_{}+{\Pmagt}_{}+{\Pmago}_{}$, (b) total
``feedback'' pressure ${\Pfb}\equiv{\Pturb}_{}+{\Pth}_{}+{\Pmagt}_{}$, and (c)
kinetic pressure ${\Pturb}_{}+{\Pth}_{}$.  We also plot as a solid line the
fitting result from HD models in Paper~I for a wide range of $\Sigma$ and
$\rhosd$, $\SigSFRnorm=2.1[{(\Pturb+\Pth)}/10^4\kbol\Punit)]^{1.18}$.  The
vertical dotted line in panel (a) denotes the total ISM weight when $\chi=1$
(as in Table~\ref{tbl:mean}), $\PDE=\mathcal{W}_{0}=\pi G \Sigma^2(1+\chi)/2
\rightarrow 6.6\times10^3\kbol\Punit$.  As shown in Figure~\ref{fig:balance},
vertical dynamical equilibrium indeed holds so that the total midplane pressure
adjusts to match this weight.  We note that the total midplane pressure and
weight are very similar for all models ($\chi\sim1$ agrees within 25\% for all
cases), since we fix $\Sigma$ and $\rhosd$, and $\sigmaeff$ and $\zeta_d$ are
insensitive to the model parameters. The vertically aligned symbols in
Figure~\ref{fig:sfr}(a) show graphically the consistency of ${\Ptot}_{,0}$ and
$\mathcal W_{0}$ in all models, and also make clear that the inclusion of
magnetic terms reduces $\SigSFR$ for the same ${\Ptot}_{,0}$. 

Since the mean magnetic field can supply vertical support without star
formation (at least on timescales comparable to the disk's vertical crossing
time), the total support requirement from star formation feedback for the
vertical equilibrium can be reduced to ${\Pfb}=\PDE-{\Delta\Pmago}$ (where we
assume $\Pmagt(\zmax) \rightarrow 0$, as turbulent fields are driven only
within the star-forming layer, but we keep $\Pmago(\zmax)$ for generality,
allowing for the possibility that the mean fields are carried into the halo by
winds).  Thus, it is interesting to plot $\SigSFR$ as a function of midplane
values of ${\Pfb}$ instead of ${\Ptot}$, which we show in
Figure~\ref{fig:sfr}(b).  In the presence of magnetic fields, even with very
weak mean fields, star formation feedback can generate turbulent magnetic
fields at a significant level in a very short time.  The additional support
from turbulent magnetic fields enhances overall feedback yield with
$\etamagt\sim1-2$ on top of $\etaturb+\etath\sim 4-5$. Thus, the feedback
pressure is larger by 20\%-40\% at a given SFR for a magnetized medium compared
to an unmagnetized medium.  The dashed line in panel (b) represents
$\SigSFRnorm={\Pfb}_{,3}/(\etaturb+\etath+\etamagt)$ with
$\etaturb+\etath+\etamagt=6.6$, the feedback yield for Model MB10.  As
expected, the symbols and the dashed line in Figure~\ref{fig:sfr}(b) for
magnetized models fall systematically below (or to the right) of the solid
line, which was based on simulation results from HD models with lower total
$\eta$ than MHD models.  

Figure~\ref{fig:sfr}(c) shows another important property of  self-regulated
star-forming ISM disks.  Since $\etaturb$ and $\etath$ are nearly unchanged
with and without magnetic fields, the same relationship between kinetic
midplane pressure ${\Pturb+\Pth}$ and SFR holds as we found in Paper~I (solid
line).  However, the present work considers  only a single value of $\Sigma$
and $\rhosd$, and simulations for a wider range of these parameters, for
magnetized models, would be needed to confirm the robustness of this
conclusion.  

In short, the presence of the magnetic fields will reduce the SFR by increasing
the total efficiency of feedback ($\etaturb+\etath+\etamagt$) and reducing the
required dynamical equilibrium pressure from feedback ($\PDE-\Delta{\Pmago}$).
The general relationship between the SFR surface density and the midplane
pressure including magnetic terms can then be written as
\begin{equation}\label{eq:sfrP}
\SigSFR =10^{-3}\sfrunit\rbrackets{\frac{\etaturb+\etath+\etamagt}{5}}^{-1}
\rbrackets{\frac{\PDE-{\Delta\Pmago}}{5\times10^3\kbol\Punit}}.
\end{equation}
If $\Delta\Pmago$ is asymptotically proportional to $\SigSFR$ (which would be
true if the mean-field dynamo arises from feedback-driven fluctuations in the
magnetic and velocity fields, similar to the small-scale turbulent dynamo), a
relationship of the form $\SigSFR \propto \PDE$ with larger total efficiency of
feedback would also hold.  Alternatively, $\Delta \Pmago$ may be negligible
compared to $\PDE$ if the scale height of the mean magnetic field is large
compared to that of the neutral ISM.

\section{Summary and Discussion}\label{sec:summary}
In Paper~I, we developed realistic galactic disk models with cooling and
heating, self-gravity, sheared galactic rotation, and self-consistent star
formation feedback using direct momentum injection by SNe and time-varying
heating rates.  Here, we extend the fiducial Solar-neighborhood-like model of
Paper~I to include magnetic fields with varying initial field strengths and
distributions. We also alter the initial and boundary conditions from the
original model to minimize artificial initial vertical oscillations.  We show
in Section~\ref{sec:evol} that the overall evolution remains the same in HD
models irrespective of differences in initial and boundary conditions,
confirming the robustness of results in Paper~I. In both HD and MHD models, the
time evolution of ISM properties reaches a saturated equilibrium state within
$\sim \torb$, except for the mean magnetic field, which continues to secularly
evolve over several $\torb$ (see Figures~\ref{fig:evol1} and \ref{fig:evol2}).
In what follows, we summarize the main results  derived from analyzing the time
evolution of mean ISM properties and final saturated-state vertical profiles,
with careful consideration of the slow variation of mean magnetic fields.

\begin{enumerate}
\item \emph{Generation and Saturation of Magnetic Fields} -- Turbulent motions
in our simulations quickly develop and saturate.  The same is true for
turbulent magnetic fields  (see Figure~\ref{fig:evol2}(b)).  Beyond $\sim
\torb$, saturated levels of turbulent kinetic and magnetic energies are similar
for all MHD models except Model MA100, which has a very weak initial field.
Model MA100 converges to the same levels of $\Ekin$ and $\Emagt$ as other
models by $\sim 3 \torb$.  The final saturated state has mass-weighted mean
$\vrms{\delta {v}}\approx 5\kms $ and $\vrms{\delta {v}_A}\approx 3\kms$ for
all MHD models, corresponding to $\Ekin/\Emagt\sim 2.5$.

The growth of magnetic fields in turbulent flows is believed to be a
consequence of turbulent dynamo action, which generally refers to mechanisms of
energy conversion from kinetic to magnetic.  Dynamos are classified into
large-scale (mean field) dynamos and small-scale (fluctuation or turbulent)
dynamos. The former and latter respectively generate magnetic fields with
scales larger and smaller than the turbulence injection scale.  In many spiral
galaxies, there are significant large-scale, regular azimuthal magnetic fields
\citep[e.g.,][]{2001SSRv...99..243B}, which may result from the so-called
$\alpha-\Omega$ dynamo driven by turbulence and differential rotation
\citep[e.g.,][]{1996ARA&A..34..155B}.  Our initial conditions with mean fields
in the $\yhat$ direction is motivated by observed preferentially azimuthal (or
spiral-arm aligned) fields.  Self-consistent growth of mean fields from the
very weak primordial magnetic fields is an important and active research area
\citep[see][for recent simulations with
SN]{2008A&A...486L..35G,2013MNRAS.430L..40G}.  Although growth of mean magnetic
fields is not the main focus of this paper, some hints can be found in the
evolution of mean fields in cases with initially weak fields (Models MA100 and
MA10 in Figure~\ref{fig:evol2}(d)). 

More straightforward results from our simulations are the saturation properties
of turbulent magnetic fields, which can be  understood in the context of
small-scale dynamos.  It is broadly known that turbulent conductive flows can
amplify their own magnetic fields  through random field line stretching,
twisting, and folding \citep[e.g.,][]{1983flma....3.....Z,2005PhR...417....1B}.
Recently, \citet{2009ApJ...693.1449C} have quantified three stages of evolution
using a comprehensive set of nonlinear simulations for incompressible MHD
turbulence \citep[see also][]{2007mhet.book...85S,2012PhRvL.108c5002B}.  The
magnetic energy grows exponentially until it reaches equipartition at the
kinetic energy dissipation scale. Then, growth becomes linear until the entire
energy spectrum reaches equipartition up to the energy injection scale.  The
saturation amplitude of turbulent magnetic energy for small-scale
incompressible dynamos approaches $\sim 40\%$ of the total energy at large
Reynolds number \citep{2009ApJ...693.1449C}.  \citet{2004PhRvE..70a6308H}
obtained $\sim 30\%$ of the total energy from their compressible simulations.
\citet{2009ApJ...691.1092L} obtained similar results for various Mach numbers
from isothermal, compressible MHD simulations \citep[see
also][]{2008Sci...320..909R}.  Nearly irrespective of the Mach number, the
ratios of magnetic energy to total (kinetic+magnetic) energy in fluctuations
are $30-40\%$.\footnote{ \citet{2011PhRvL.107k4504F} investigated the growth of
magnetic energy in forced turbulence from very weak initial fields with a
variety of Mach numbers and two different forcing schemes (solenoidal and
compressive).  Their ratios of magnetic to kinetic energy at saturation are
0.4-0.6 for subsonic, solenoidally driven turbulence but only 1\% or less for
supersonic turbulence.  However, their saturated magnetic energy is not
measured at final saturation but the end of the exponential growth stage.
Based on our results (slow convergence of turbulent magnetic energy in Model
MA100), the low saturation level in the supersonic (or transonic) cases they
report may also owe to their extremely weak initial mean fields.} With stronger
initial field, saturation is achieved more rapidly.

The small-scale dynamo provides a fast and universal mechanism for magnetic
field amplification \citep{2005PhR...417....1B,2015ASSL..407..163B}.  Thus, we
expect that qualitatively similar processes will occur  in our simulations in
spite of their greater physical complexity, including compressibility, vertical
stratification, SN driven turbulence, self-gravity, and multiphase gas with
cooling and heating.  In early phases $t<\torb$, rapid growth and saturation of
the turbulent magnetic energy are induced by driven turbulence (see
Figure~\ref{fig:evol2}).  After $\torb$, turbulence is driven entirely by SN
feedback, with a rate that depends on the SFR.  Vertical dynamical equilibrium
(see below) dictates the turbulent pressure level, and hence SFR, that is
self-consistently reached in these disk systems.  The final saturated-state
turbulent magnetic energy level is about half of turbulent kinetic energy, as
expected in small-scale dynamo simulations.

\item \emph{Vertical Structure of the Diffuse ISM} --
The vertical distribution and gravitational support of the diffuse ISM in the
Milky Way has been investigated many times under the assumption of vertical
dynamical equilibrium\footnote{We prefer to use the term ``vertical dynamical
equilibrium'' instead of ``hydrostatic equilibrium'' since the gas is not
``static'' -- turbulent pressure arising from large-scale gas motions is
crucial to the vertical force balance.}
\citep{1990ApJ...365..544B,1991ApJ...382..182L,1998A&A...339..745K}.  In
Section~\ref{sec:equil}, we validate that vertical dynamical equilibrium is
satisfied in turbulent, star-forming, magnetized galactic disks, extending
results from previous multiphase ISM  simulations with various physical
ingredients
\citep{2007ApJ...663..183P,2009ApJ...693.1346K,2010ApJ...720.1454K,2011ApJ...743...25K,2013ApJ...776....1K,2012ApJ...750..104H}.
At the midplane, vertical support is dominated by the turbulent (kinetic)
pressure except in model (MB1) with very strong mean magnetic fields (see
Figure~\ref{fig:balance} and Table~\ref{tbl:meanP}).  Thermal and turbulent
magnetic supports are comparable to each other and about 2-3 times smaller than
the turbulent pressure.  Compared to HD models, MHD disk models can maintain
smaller turbulent and thermal pressures owing to additional vertical support
from the turbulent and mean magnetic fields.  Despite of different
magnetization, the total effective velocity dispersion $\sigmaeff$ varies by
only $\sim 25\%$ over all models (see Table~\ref{tbl:meanv}), resulting in
similar variation in $\chi$ and hence dynamical equilibrium midplane pressure
$\PDE=\mathcal{W}_0={\Wsg}_{,0}(1+\chi)$.

Our models demonstrate that vertical dynamical equilibrium is indeed a good
assumption even though disks are magnetized and highly dynamic.  However,
practical application of vertical equilibrium in observations is not simple.
Even in the Solar neighborhood, uncertainty is substantial.  For example,
\citet{1990ApJ...365..544B} calculated the total equilibrium pressure based on
observations of the vertical gas distribution and gravity
\citep[e.g.,][]{1987A&A...180...94B,1989MNRAS.239..651K}, and taking the
difference with observed thermal+turbulent pressure inferred significant
contributions from non-thermal (cosmic-ray and magnetic) pressures.  However,
\citet{1991ApJ...382..182L} have argued that local (high-latitude) \ion{H}{1}
21 cm emission data can be fitted with three-component Gaussians, and that
non-thermal vertical support is not necessary to explain the \ion{H}{1}
distribution.  Considering the large observed scale height of non-thermal
pressure terms,  the vertical support of \ion{H}{1} within $|z|<1\kpc$ (which
depends on the pressure gradient, not pressure itself), can be explained solely
by thermal + turbulent terms.  Later, \citet{1998A&A...339..745K}
simultaneously analysed the distribution of emission at 21 cm from warm/cold
\ion{H}{1} with H$\alpha$ from diffuse ionized gas, soft X-rays for the hot
medium, and synchrotron radiation combined with $\gamma$-ray emission probing
magnetic fields and cosmic rays.  They concluded that within $400\pc$ of the
midplane, cosmic ray support is not required, and turbulent magnetic fields
contribute with $\mathcal{R}\sim 1/3$ ($\alpha=1/3$ in their notation).  The
contribution of turbulent magnetic support compared to thermal+turbulent
kinetic pressure in our simulations is $\sim 25\%$ from Table~\ref{tbl:meanP},
in good agreement with this result.  In our simulations, the mean magnetic
field has a relatively small scale height (Figure~\ref{fig:Pprof}) and vertical
support slightly less than that of the turbulent magnetic field.  In contrast,
observations indicate that magnetic fields extend into the halo in equipartion
with the pressures of cosmic rays and hot gas.  The decline in mean magnetic
fields with height in our present simulations may be due to the absence of a
hot medium and cosmic rays; this issue will be addressed in future work.

Adoption of vertical equilibrium is useful to estimate the midplane pressure in
external galaxies.  The pressure cannot be measured directly except for nearby
edge-on disks, and even in that case requires deprojection and an assumption
for the turbulent velocity dispersion
\citep{2011AJ....141...48Y,2014AJ....148..127Y}. As outlined in
Section~\ref{sec:theory}, to get the total dynamical equilibrium pressure (or
the total weight of the gas; see Equation (\ref{eq:de})), one may need to
determine $\Sigma$, $\Sigma_*$, $\sigma_*$, and $\sigmaeff$ (or $\sigma_z$ and
$\mathcal{R}$), assuming the parameter $\zeta_d\sim 0.4-0.5$ is nearly
constant.  In several systematic studies of nearby galaxies
\citep{2004ApJ...612L..29B,2006ApJ...650..933B,2008AJ....136.2782L,2008AstL...34..152K},
$\sigmaeff$ and/or $H_*\propto \sigma_*^2/\Sigma_*$ are assumed to be constant
to determine the total midplane pressure.  Observational measurements have
suggested that \ion{H}{1} velocity dispersions are nearly constant with
$\sigma_z\sim 5-10\kms$, especially for the atomic gas dominated regime, using
single Gaussian fitting \citep{2007AJ....134.1952P} and intensity-weighted
second velocity moments  \citep{2009AJ....137.4424T}.  Recent studies of global
\ion{H}{1} kinematics analysis to obtain ``superprofiles'' (averaged \ion{H}{1}
profiles for the entire galaxy) in nearby spiral and dwarf galaxies have
arrived at similar conclusions \citep{2013ApJ...765..136S,2012AJ....144...96I}.
In this paper, we show that the contribution from turbulent magnetic field to
$\mathcal{R}$  (or, equivalently, $[\delta {v}_A^2/2 - \delta
v_{A,z}^2]/\sigma_z^2)$ is $\approx 0.3$ insensitive to the mean magnetic field
strength for saturated models (except Model MB1).  This suggests that if the
total velocity dispersion $\sigma_z$ can be measured accurately, an estimate of
$\sigmaeff$ needed to compute $C$ and $\chi$ and therefore the total midplane
pressure may be obtained using a typical value of $\mathcal{R}\sim0.3$ even if
the observed magnetic field strength is not measured.  

The density structure in our simulations is well characterized by two
components, a cold and non-cold medium (see Figure~\ref{fig:density}).  For the
present models, we find scale heights of cold and non-cold components are
$H\sim30\pc$ and $110\pc$, respectively. \citet{1991ApJ...382..182L} have used
three-component Gaussians to fit \ion{H}{1} emission line observations near the
Sun, obtaining vertical scale heights equivalent to $H\sim80-130\pc$, $\sim
150-300\pc$, and $\sim 600-750\kpc$ \citep[see also][]{1990ARA&A..28..215D}.  A
recent {\it Herschel} galactic survey of [\ion{C}{2}],  along with ancillary
\ion{H}{1}, ${}^{12}$CO, ${}^{13}$CO, and C${}^{18}$O data, shows that the
equivalent vertical scale heights of [\ion{C}{2}] sources with CO and without
CO which trace colder/denser vs. warmer/more diffuse gas, respectively, are
$\sim70\pc$ and $170\pc$ \citep{2014A&A...564A.101L,2014A&A...572A..45V}.  Our
simulations agree with both of these observational studies in the sense that
the warmer component has $\sim 2-3$ times the scale height of the colder
component.  However, our measured $H$ values are somewhat lower than observed
Solar-neighborhood values, as our measured values of $\sigma_z$ and $\SigSFR$.
Simulations currently underway that include a hot ISM suggest that $H$,
$\sigma_z$, and $\SigSFR$ may increase.

\item \emph{Feedback Efficiencies and Star Formation Laws} --
One of the important conclusions of this study is that the feedback yields for
turbulent and thermal pressure are unchanged by the presence of magnetic
fields.  This is mainly because the HD and MHD turbulent energy dissipation
rates are similar. The dissipation time scale is always of order the  crossing
time at the driving scale (or main energy-containing scale) both for HD
turbulence \citep{2003PhFl...15L..21K} and MHD turbulence
\citep{1998ApJ...508L..99S,1998PhRvL..80.2754M,2009ApJ...691.1092L,2009ApJ...693.1449C},
and both compressible and incompressible flows.  Turbulent magnetic fields
saturate at a similar timescale to turbulent velocities.  Balancing turbulent
driving with turbulent dissipation therefore leads to direct proportionality
between $\SigSFR$ and the turbulent kinetic and magnetic pressures.  Similarly,
balancing heating and cooling leads to a direct proportionality between
$\SigSFR$ and the thermal pressure.  turbulent kinetic and magnetic pressures.
Defining ``yield'' as the pressure-to-$\SigSFR$ ratio (in convenient units; see
Equation~\ref{eq:eta}), we obtain turbulent, thermal, and turbulent magnetic
feedback yields as $\etaturb\sim 3.5-4$, $\etath\sim 1.1-1.4$, and
$\etamagt\sim 1.3-1.5$, respectively.  Since the ISM weight and therefore total
pressure is nearly the  same irrespective of magnetization, the addition of
magnetic terms to the total $\eta$ reduces $\SigSFR$ for MHD compared to HD
simulations at a given $\Sigma$ and $\rhosd$.

Correlations between the estimated ISM equilibrum pressure $\PDE$ and the
molecular content and star formation rate have been identified empirically
\citep[e.g.,][]{2002ApJ...569..157W,2006ApJ...650..933B,2008AJ....136.2782L}.
In particular, the $P-\SigSFR$ relation has much less scatter than the
classical Kennicutt-Schmidt relationship \citep[][]{2012ARA&A..50..531K}
between gas surface density $\Sigma$ and $\SigSFR$ for the atomic-dominated
regime \citep[e.g.][]{2008AJ....136.2846B,2010AJ....140.1194B}.  As argued in
our previous work
\citep[e.g.,][Paper~I]{2010ApJ...721..975O,2011ApJ...743...25K}, this is
because the total midplane pressure is directly related to the SFR, whereas ISM
properties (and the SFR) can vary considerably at a given value of $\Sigma$
depending on the gravitational potential confining the disk.  As shown in
\citet[][see also \citealt{2012ApJ...754....2S}]{2011ApJ...731...41O}, in the
starburst regime where self-gravity dominates the potential, both pressure and
surface density correlate well with SFR surface density, giving
$\SigSFR\propto\PDE\propto\Sigma^2$ (shallower reported slopes are arguably due
to too-high assumed CO-to-H$_2$ ratios at high $\Sigma$; see
\citealt{2012MNRAS.421.3127N}). In the atomic-dominated regime, as modelled in
this paper, the dynamical equilibrium pressure depends on both
$\mathcal{W}_{\rm sg}\propto\Sigma^2$ and $\mathcal{W}_{\rm ext}\propto
\Sigma\sqrt{\rhosd}$ (see Equations (\ref{eq:wsg}) and (\ref{eq:wext})).  In
outer disks where $\rhosd$ varies more than $\Sigma_{\mathrm{atomic}}\sim
6-10\Msun \pc^{-2}$,  simple Kennicutt-Schmidt laws fail. 

As demonstrated in this paper, there can also be scatter introduced in
$\SigSFR$ vs.  $\PDE$ if support from the mean magnetic field contributes
substantially without itself having $\Delta\Pmago \propto \SigSFR$.  Even if a
mean-field dynamo secularly leads to a well-define asymptotic ratio between
$\Delta\Pmago$ and $\SigSFR$, the long timescale to reach this may mean that
$\Delta\Pmago$ is not well correlated with the recent SFR.  If the vertical
scale height of the mean field is larger than that of the star-forming gas,
however, the support from $\Delta\Pmago$ would be small even if the mean field
pressure is non-negligible, which would reduce scatter in $\SigSFR$ vs. $\PDE$.

\end{enumerate}

\acknowledgments
This work was supported by grants AST-1312006 from the National Science
Foundation and NNX14AB49G from NASA.  We thank E. Blackman and J.-M. Shi for
discussions of dynamos, and the referee for helpful comments on the
manuscript.

\appendix
\section{Numerical Convergence}

We adopt a standard spatial resolution of $\Delta=$2pc throughout this paper. 
Here, we present a convergence study for physical properties at the saturated
state.  Since a simulation with $\Delta=$1pc for the same box size would be
too computationally expensive for our current facilities, we instead have rerun
Model MB10 with $\Delta=$1pc, 2pc, and 4pc using a smaller horizontal box
($L_x=L_y=256\pc$) and a halved the turbulence driving period ($t_{\rm
drive}=0.5\torb$) and final time ($t_{\rm end}=2\torb$).

\begin{figure}
\plotone{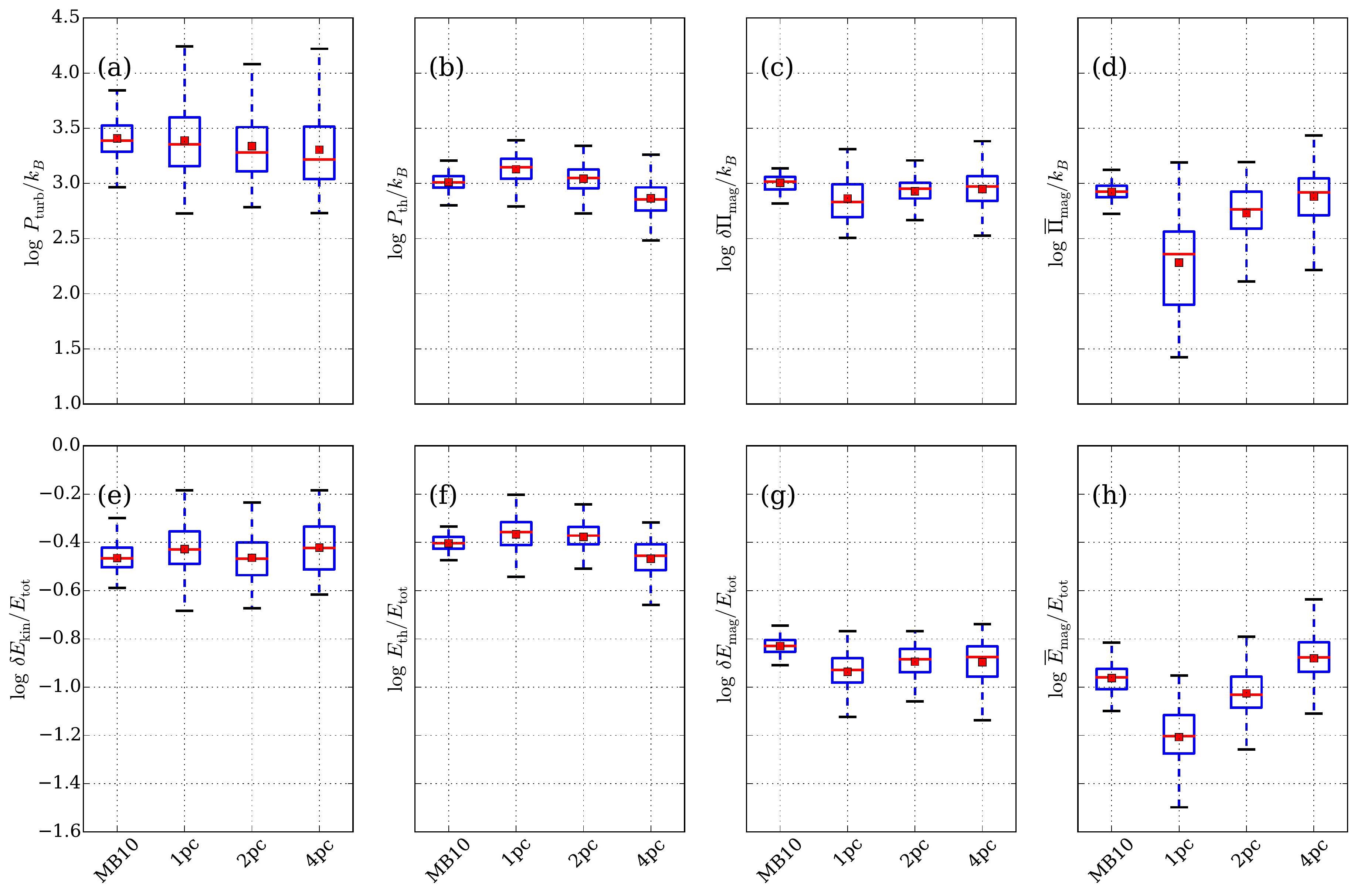}
\caption{
Box and whisker plots of the logarithms of the midplane support components
({\em top}; (a)-(d)) and the energy ratios ({\em bottom}; (e)-(h)) over the
interval $(\torb,2\torb)$.  From {\em left} to {\em right} columns, we show the
turbulent kinetic, thermal, turbulent magnetic, and mean magnetic components.
We label the original MB10 model as `MB10', while the smaller box counterparts
at three different resolutions are labeled `1pc', `2pc', and `4pc'.}
\label{fig:convergence}
\end{figure}

Figure~\ref{fig:convergence} summarizes the midplane support components and the
energy ratios for $(\torb, 2\torb)$ with box-and-whisker plots. We plot the
quantities by taking the logarithms for clearer presentation of temporal
fluctuations.  We label the original MB10 model (full box with $\Delta=2pc$)
as `MB10' and the smaller box counterparts at three different resolutions
as `1pc', `2pc', and `4pc'.  Although the smaller box increases temporal
fluctuations as expected, all components except the mean magnetic component
show statistically converged results.

The mean magnetic component in the 1pc simulation is systematically reduced,
and has the largest temporal fluctuation, also causing a slightly smaller
saturated value of the turbulent magnetic component. Considering that the mean
magnetic component always shows the slowest convergence, it is possible that
with a longer integration time, the mean magnetic field in the 1pc model would
approach that of the other models.  However, it is also possible that it would
remain at this reduced level. Since the evolution of mean magnetic fields is
not the main focus of this paper, we defer further study of the mean magnetic
component to future work, which will include more comprehensive investigation
of mean field dynamo.

\end{document}